\documentclass[a4paper,10pt]{article}

\usepackage{lscape}
\usepackage{cancel}
\usepackage{bm}
\usepackage{bbm}
\usepackage{multirow} 
\usepackage{amsfonts}
\usepackage{amssymb}
\usepackage{graphicx}
\usepackage{amsmath}
\usepackage[usenames,dvipsnames]{color}
\usepackage{hyperref}
\usepackage{epstopdf,epsfig}
\usepackage{verbatim}
\usepackage[utf8]{inputenc}
\usepackage[left=1in,right=1in,top=1in,bottom=1in]{geometry}             

\hypersetup{
    colorlinks=true,         
    linkcolor=blue,          
    citecolor=red,        
    urlcolor=Violet             
}

\newcommand{\order}[1]{\ensuremath{\mathcal{O}(#1)}}

\def\be{\begin{equation}}
\def\ee{\end{equation}}
\def\bea{\begin{eqnarray}}
\def\eea{\end{eqnarray}}

\title{Extending the rigidity of general relativity}
\author{Henrique Gomes and Vasudev Shyam}
\author{{\bf Henrique Gomes\footnote{\href{mailto:gomes.ha@gmail.com}{gomes.ha@gmail.com}}}\, and {\bf Vasudev Shyam}\footnote{\href{mailto:vshyam@pitp.ca}{vshyam@pitp.ca}}
\\\it Perimeter Institute for Theoretical Physics\\ \it 31 N. Caroline St. Waterloo, ON, N2L 2Y5, Canada }
\begin{document}
\maketitle
\begin{abstract}
We give the most general conditions to date which lead to uniqueness of the general relativistic Hamiltonian. Namely, we show that all spatially covariant generalizations of the scalar constraint which extend the standard one while remaining quadratic in the momenta are second class. Unlike previous investigations along these lines, we do not require a specific Poisson bracket algebra, and the quadratic dependence on the momenta is completely general, with an arbitrary local operator as the kinetic term.   
\end{abstract}

\section{Introduction}

The key feature of (metric) General Relativity is that it equates
relativistic gravitation to a purely geometric theory of spacetime
where the dynamical variable is the space-time metric tensor $^{(4)}g_{\mu\nu}$
and the dynamical law is a second order partial differential equations
for the metric. These  are know known as the Einstein field equations (written in
natural units):
\begin{equation}\label{equ:Einstein}
^{(4)}R_{\mu\nu}-\frac{1}{2}{}^{(4)}g_{\mu\nu}{}^{(4)}R=T_{\mu\nu}.
\end{equation}
The field equations relate the curvature of spacetime
to the energy momentum tensor lying on the rhs of \eqref{equ:Einstein}. The energy momentum tensor, $T_{\mu\nu}$, describes the matter content
of the theory. We shall mostly deal with the vacuum field equations
in this work and shall set $T_{\mu\nu}=0$. In four dimensions this
theory still has non-trivial dynamical local degrees of freedom.\footnote{ This fact can most easily be seen at the linearized
level, where the vacuum Einstein equations describe the propagation of massless spin-2 gravitons.  For the non-perturbative statement, we use the canonical theory, in the main text.}

 For most of the paper, we will not be dealing with the metric of space-time, but with the Riemannian metric of spatial hypersurfaces. For this reason, we use the superscripts $^{(4)}$ to denote quantities related to space-time, as opposed to  spacelike
hypersurface quantities,  for which we shall employ no
superscripts. 

 The use of such Riemannian metrics and spatial hypersurfaces are required for the treatment of  relativistic
gravitation as a Hamiltonian dynamical system, known as  Canonical General Relativity. This formalism is due to the work of Arnowitt, Deser and Misner in \cite{ADM} and is thus also known as the ADM formalism. The phase space variables are the metric on a space-like hypersurface and its conjugate momentum:
$(g_{ab}(x),\pi^{ab}(x)),$ (the explicit dependence on space via
the label $x$  reminds us that this is the pair of canonically
conjugate variables per space point in an infinite dimensional phase
space). So, at each point, we have 12 phase space dimensions as both
the metric and its conjugate momentum are symmetric tensors. This is the counting at the kinematic level, before the imposition of constraints.

The Hamiltonian
of General Relativity is a sum of four constraints, the scalar Hamiltonian
constraint (one per point) and the vector difffeomorphism constraint (three per point). These constraints,
together with their Poisson algebra, encapsulate the initial value problem, the gauge symmetries and the evolution of the theory. 

At the full non linear level, one can find the non-triviality of the local degrees of freedom of general relativity by the usual counting for Hamiltonian constrained systems: 
 take  the difference between the
number of dimensions of the phase space and twice the number of first
class constraints (added to the number of second class constraints should
there be any), and divide by two. We have four first class constraints and the above
arithmetic yields 2 degrees of freedom per space point. At the linearized
level, these correspond to the polarization states of the graviton. Going back
to the key feature of General Relativity being the geometric nature
of its dynamical law, we should expect it to manifest also 
in the Hamiltonian version of the theory, and indeed it does.

   The Hamiltonian, or
scalar, constraint reads:
\begin{equation}
H(N)=\int\textrm{d}^{3}x\ N(x)\left(\frac{G_{abcd}\pi^{ab}\pi^{cd}}{\sqrt{g}}-\sqrt{g}(R-2\Lambda)\right)=0,
\end{equation}
 where $G_{abcd}=\left(g_{ac}g_{bd}+g_{ad}g_{bc}-\frac{1}{2}g_{ab}g_{cd}\right)$
is known as the De-Witt supermetric, and the lapse function $N(x)$
is the Lagrange multplier enforcing this constraint. The momentum
or diffeomorphism constraint reads

\begin{equation}
H_{a}(N^{a})=-\int\textrm{d}^{3}x\ N^{a}(x)\left(2\nabla_{b}\pi_{a}^{b}\right)=0.
\end{equation}
 Here, the shift vetor $N^{a}(x)$ is the Lagrange multiplier enforcing
the momentum constraint. The total Hamiltonian is just the sum $\mathcal{H}_{tot}=H(N)+H_{a}(N^{a}).$
We will denote the canonical Poisson brackets as $\{\cdot,\cdot\}$,
and the Poisson bracket between the fundamental phase space conjugate
variables $(g_{ab}(x),\pi^{ab}(x))$ reads 

\[
\{g_{ab}(x),\pi^{cd}(y)\}=\delta_{a}^{(c}\delta_{b}^{d)}\delta(x-y).
\]
 Note that by $\delta(x-y)$ we really mean $\delta^{(3)}(x-y).$
Thus the action of the Poisson brackets on arbitrary Phase space functions
$F(g(x),\pi(x))$ and $P(g(y),\pi(y))$ is given by:

\[
\{F(g(x),\pi(x)),P(g(y),\pi(y))\}=\int\textrm{d}^{3}z\left(\frac{\delta F(x)}{\delta g_{ab}(z)}\frac{\delta P(y)}{\delta\pi^{ab}(z)}-\frac{\delta P(y)}{\delta g_{ab}(z)}\frac{\delta F(x)}{\delta\pi^{ab}(z)}\right).
\]
 Using this, we can also compute the Poisson algebra of the constraints,
which is known as the Dirac or hypersurface deformation algebra (and
we will discuss why this is so). The algebra is:

\begin{equation}\label{equ:scalar_algebra}
\{H(N),H(M)\}=H_{a}(g^{ab}\left(N\nabla_{b}M-M\nabla_{b}N\right)),
\end{equation}

\begin{equation}\label{equ:momentum_algebra}
\{H(N),H_{a}(M^{a})\}=H(\mathcal{L}_{M^{a}}N),
\end{equation}

\begin{equation}
\{H_{a}(N^{a}),H_{b}(N^{b})\}=H_{c}(N^{a}\nabla_{a}M^{c}-M^{a}\nabla_{a}N^{c}).
\end{equation}
 The explicit dependence on the metric tensor in the first equation
above indicates that this Poisson algebra strictly speaking isn't
a lie algebra because the structure functions depend on the canonical
variables too. 

As mentioned above, given the geometric nature of the dynamical law in
General Relativity,  this Poisson algebra inherits a geometric interpretation --
 it reflects the algebra of deformations of a hypersurface
embedded into space-time. In other words, decomposing vectors into parallel and orthogonal components along a hypersurface, one obtains an algebra from their commutators as space-time vector fields. This abstract algebra is the same as that obeyed by the constraints with the Poisson bracket above \eqref{equ:scalar_algebra}, \eqref{equ:momentum_algebra}, as shown in the seminal paper, \cite{HKT}.

Their result implies that  generators of the algebra encode the possible deformations of the hypersurface as embedded in space-time. 
In this restricted sense, the structure of the Poisson algebra of the constraints
of canonical gravity encodes space-time diffeomorphism invariance. 
Although the encoding of full general covariance in the canonical theory is
indirect, one could argue  that this indirectness is to be expected: in doing
the canonical analysis, manifest covariance needs to be  sacrificed, as
a distinction is to be made between quantities intrinsic and extrinsic
to the space(or time)like hypersurface.

 Intrinsically, the diffeomorphism
constraint encodes covariance on the hypersurface quite straightforwardly.
This can be seen by noting that we can infinitesimally Lie-drag a
tensor on phase space $F_{i_{1}\cdots i_{k}}^{j_{1}\cdots j_{l}}[g,\pi,x)$
along the flow generated by a vector field $\xi^{a}$ by computing
the following Poisson bracket:

\[
\{H_{a}(\xi^{a}),F_{i_{1}\cdots i_{k}}^{j_{1}\cdots j_{l}}[g,\pi,x)\}=\mathcal{L}_{\xi}F_{i_{1}\cdots i_{k}}^{j_{1}\cdots j_{l}}[g,\pi,x).
\]

The interpretation of the scalar constraint is perhaps less straightforward. It is only related to non-spatial space-time diffeomorphisms if the Einstein equations of motion are satisfied \cite{Wald_Lee}. Nonetheless,  its commutation algebra with the diffeomorphism
constraint in the above form acknowledges that the hypersurface is embedded in an ambient
space-time. 

This rigid relation between the constraints, their algebra, and geometric features of  space-time, suggests  that locally altering the structure of the constraints
should lead to the sacrifice of  some of these features; including general covariance, depending
on the nature of the modification.
Of such modifications, those that are local and preserve spatial covariance but
 risk sacrificing space-time covariance are of particular interest -- studying them might allow us to pin-point which features of
the constraints and their structure are essential for general covariance
and which are superfluous. 

Furthermore, sacrificing space-time covariance
and retaining solely spatial covariance changes the number of propagating degrees
of freedom from two to three. Thus, a secondary question would be: what deformations
or modifications of the constraints would  still preserve
the number of propagating degrees of freedom? By this we mean those that leave the Hamiltonian constraint first class. If there are any such modifications, then they describe
theories of gravity where general covariance in the usual sense is
lost but the number of degrees of freedom remain the same as in general
relativity. If we find that no such modifications are possible, then
we can better understand the uniqueness of general relativity
in describing relativistic gravitation.

\section{Modifying the Hamiltonian Constraint }

As mentioned in the previous section, the class of modification
we are considering are those that alter the Hamiltonian constraint
while respecting spatial covariance. First, we recall that the Hamiltonian
constraint 

\[
H(N)=\int\textrm{d}^{3}x\ N(x)\left(\frac{G_{abcd}\pi^{ab}\pi^{cd}}{\sqrt{g}}+\sqrt{g}(R-2\Lambda)\right)
\]
 consists of a `kinetic energy' term $F_{o}[g,\pi;x)=\frac{G_{abcd}\pi^{ab}\pi^{cd}}{\sqrt{g}}(x)$
and a `potential energy' like term $V_{o}[g;x)=\sqrt{g}(R-2\Lambda)(x).$
The modifications we consider are of the form

\be\label{simple_expansion}
F_{o}[g,\pi;x)\rightarrow F_{o}[g,\pi;x)+F[g,\pi;x)\equiv\Phi[g,\pi;x),\ V_{o}[g;x)\rightarrow V_{o}[g;x)+V[g;x)\equiv\hat{V}[g;x),
\ee
 where 
\begin{equation}
F[g,\pi;x)=\sum_{r,m,n}\mathcal{B}_{abcd}^{i_{1}\cdots i_{n}\,j_{1}\cdots j_{m}}\left(\nabla_{i_{1}}\cdots\nabla_{i_{n}}\pi^{cd}\right)\left(\nabla_{j_{1}}\cdots\nabla_{j_{m}}\pi^{ab}\right)(x)\label{eq:kinetic_original}
\end{equation}
 here the tensor $\mathcal{B}_{abcd}^{i_{1}\cdots i_{n}\,j_{1}\cdots j_{m}}\mathcal{}$
and the scalar function $V[g,x)$ both depend on the metric $g_{ab}$
and its derivatives through $\ensuremath{R_{kl},\nabla_{a_{1}}R_{kl},\cdots,\nabla_{a_{1}}\cdots\nabla_{a_{r}}R_{kl}}$
where the number of explicit covariant derivatives, $r$, is arbitrary. Note that the spatial derivative order of $\mathcal{B}_{abcd}^{i_{1}\cdots i_{n}\,j_{1}\cdots j_{m}}\mathcal{}$
and $V[g,x)$ differ in all generality. As explained in the previous section, we would like
to search for modifications of this kind that leave the Hamiltonian
constraint first class. 

Before describing how we propose to do so,
it is worth noting that these modifications -- which have $n+m+r+2$ spatial derivatives and 2 time derivatives -- are necessarily breaking
Lorentz invariance due to the imbalance between the number of spatial
and time derivatives. This means that if any of these modifications
maintain the first class nature of the Hamiltonian constraint, then
they will describe Lorentz non covariant gravitational theories that
propagate the same number of degrees of freedom as General Relativity.
It is also known from the work of Frackas and Martinec \cite{Martinec} that
if we were to allow  $V[g,x)$ to be non zero but restrict ourselves to the situation where $F[g,\pi,x)=0$,
i.e. to the case where the kinetic term is ultralocal in both the
momenta and the metric,  then
the demand that the Hamiltonian constraint remain first class forces
$V[g,x)$ to vanish as well.

  The first class condition is given by:
\[
\left\{ H(f),H(h)\right\} \approx0.
\]
where the $\approx$ symbol denotes weak equality, meaning equality when
the constraints of the theory are satisfied. If it turns out that we find such first class, we expect that their zero locus
on phase space is a sub-manifold that is integrable. This equation
can also be written as a `strong' equation -- valid everywhere
on phase space -- as:
\be\label{general_first_class}
\left\{ H(x),H(y)\right\} =\int\textrm{d}^{3}z\left(W_{H(x)H(y)}^{H(z)}[g,\pi;,x,y)H(z)+U_{H(x)H(y)}^{H_{a}(z)}[g,\pi;,x,y)H^{a}(z)\right).
\ee
 Our task then is to compute the structure functions $W_{H(x)H(y)}^{H(z)}$ and $ U_{H(x)H(y)}^{H_{a}(z)}$
if at all possible.\footnote{We have here performed the usual extension from the structure functions of first class constraint brackets, $\{\chi_\alpha, \chi_\beta\}=C^\gamma_{\alpha,\beta}\chi_\gamma$ to the functional context. } 

Expanding the bracket using the first simple expansion \eqref{simple_expansion}:
\[
\left\{ H(f),H(h)\right\} =\left\{ \Phi(f)+V(f),\Phi(h)+V(g)\right\} =\left\{ F_{o}(f)+F(f)+\hat{V}(f),F_{o}(f)+F(f)+\hat{V}(f)\right\} 
\]
\[
=\left\{ F_{o}(f)+F(f),F_{o}(h)+F(h)\right\} +\left\{ F_{o}(f)+F(f),\hat{V}(h)\right\} 
+\left\{ \hat{V}(f),F_{o}(h)+F(h)\right\} +\cancelto{0}{\left\{ \hat{V}(f),\hat{V}(h)\right\} }
\]

\[
=\cancelto{0}{\left\{ F_{o}(f),F_{o}(h)\right\} }+\left\{ F(f),F(h)\right\} 
+\left\{ F(f),F_{o}(h)\right\} -\left\{ F(h),F_{o}(f)\right\} 
\]

\[
+\left\{ F_{o}(f)+F(f),\hat{V}(h)\right\} +\left\{ \hat{V}(f),F_{o}(h)+F(h)\right\} .
\]
 The reason the bracket $\{\hat{V}(f),\hat{V}(h)\}$ vanishes is because
the potential is a function of the metric alone, similarly the ultra-local
nature of the super metric is responsible for the vanishing of $\left\{ F_{o}(f),F_{o}(h)\right\} $. That is, for ultra-local terms there are no derivatives of the smearing functions $f$ and $h$, and thus the anti-commutative nature of the bracket ensures its vanishing. 

Now, in order for the constraint to remain first class, each of the other
terms need to vanish when the constraints are imposed. This is because
there is a separation in the powers of the momentum $\pi^{ab}$ and
the spatial derivative order among the different terms. The Poisson brackets reduce the order of the momenta
by one and leave the order of the spatial derivatives unchanged. There
cannot be cancellation across terms of different powers of momentum
or of different order in derivatives. So the independent terms are
organized as in table \ref{table} above. 
\begin{table}\caption{Terms organized by derivative and momentum order.}
\begin{center}
\begin{tabular}{|c|c|c|}
\hline 
 & $\mathcal{O}(\partial^{(p)}g)$ & $\mathcal{O}(\partial^{(p)}g)\cdot\mathcal{O}(\partial^{(q)}g)$\tabularnewline
\hline 
$\mathcal{O}(\pi)$ & $\left\{ F_{o}(f),\hat{V}(h)\right\} +\left\{ \hat{V}(f),F_{o}(h)\right\} $ & $\left\{ F(f),\hat{V}(h)\right\} +\left\{ \hat{V}(f),F(h)\right\} $\tabularnewline
\hline 
$\mathcal{O}(\pi^{3})$ & - & -\tabularnewline
 &  & \tabularnewline
\hline 
\hline 
 & $\mathcal{O}(\partial^{(q)}g)\cdot\mathcal{O}(\partial^{(q)}g)$ & $\mathcal{O}(\partial^{(q)}g)$\tabularnewline
\hline 
$\mathcal{O}(\pi)$ & - & -\tabularnewline
\hline 
$\mathcal{O}(\pi^{3})$ & $\left\{ F(f),F(h)\right\} $ & $\left\{ F(f),F_{o}(h)\right\} -\left\{ F(h),F_{o}(f)\right\} $\tabularnewline
\hline 
\end{tabular} 
\end{center}\label{table}
\end{table}

 Here, $(p)$ denotes the highest spatial derivative order in $\hat{V}[g,x)$
and $(q)$ denotes the highest spatial derivative order in $F[g,\pi,x)$.
We emphasise again that there cannot be any interference across the
different entries of the table because of distinct orders of spatial derivative
and powers of the momentum. In the special case
where $p=q$ for instance, although the spatial derivative order of
the first entry of the first table and the last of the second are
the same (and similarly for the second entry of the first table and
the first of the second table) there still cannot be interference
or cancellation because the polynomial order in the momenta are different.
This means that each of the entries of the table must vanish on the
constraint hypersurface separately in order for the Hamiltonian constraint
to remain first class. 

 Our focus
will be the bottom right entry of the second table, namely the term 

\[
\left\{ F(f),F_{o}(h)\right\} -\left\{ F(h),F_{o}(f)\right\} .
\]
 The reason we choose this over the others is because it pertains
solely to the kinetic term of the Hamiltonian constraint -- doesn't
involve the potential at all -- and it contains less unknowns than the other term of order $\pi^3$ (last line in table \ref{table}). 

  As mentioned, the ultralocality
of the kinetic term in both the metric and the momenta is a heavily relied upon condition for the analysis of 
 Frakas and Martinec \cite{Martinec}.  To extend the analysis to more general kinetic terms, it proves useful to 
have  a criterion for the first class property expressed solely in terms of the kinetic term itself. If we can find what sub class of modifications can lead to
the vanishing of this term on the constraint surface, we can then
proceed to checking whether such a term would survive the other entries
of the table. However,  if we can show that no modification beyond the original 
ultralocal term can survive, then the $\order{\pi}\times \order{\partial^{(q)}g}$ term is all we will need to
study. In what follows we will show that the latter is the case. I.e. we will prove that no modification of the kinetic term
beyond one that is ultralocal in both the metric and the momenta,
as is the case in General Relativity, will leave the Hamiltonian constraint
first class. 

\subsection*{Comparison to earlier work: a guide for the perplexed}
 Before we delve into the details of our construction, it is helpful to contextualize our work. In the subject of rigidity of general relativity as a dynamical system, we mention two main references, \cite{HKT, Martinec}. The seminal paper \cite{HKT} obtained rigidity results assuming that the commutation algebra between constraints was that of the embedded hypersurface decomposition of vector fields (besides assumptions of locality that we also implement). They also assumed invertibility criteria between the momentum and the extrinsic curvature, which imposes further restrictions than us on the kinetic term.

  An alternative motivation for this work is to understand the emergence of covariance and what it has to do with  gauge symmetries, or more specifically, the maintenance of the number of propagating degrees of freedom of gravitational theories. It has been shown by Weinberg  that the linearised gauge symmetry of the graviton field is entirely a consequence of locality, unitarity and Lorentz invariance \cite{Weinberg}. This fact, together with the theorem of Weinberg and Witten \cite{Weinberg_Witten} which shows that a massless spin-2 particle cannot be a composite particle, uniquely specifies linearised general relativity as the only consistent theory of interacting massless gravitons. However, these arguments of uniqueness can\rq{}t be extrapolated to scenarios where one demands the maintenance of the number of physical degrees of freedom but is willing to sacrifice Lorentz invariance.
 
  So a valid question to ask is whether there is sense in which Lorentz invariance emerges, and this was the motivation behind work of  of Khoury et al, \cite{Khoury}. They worked in an effective field theory setting assuming only spatial diffeomorphism invariance and showed that the first non trivial terms in the dynamical Hamiltonian which propagate the spatial diffeomorphism constraint are those of general relativity. 
  Although we don't set up the problem in quite the same way as they do, we too demand maintaining the same number of physical degrees of freedom as in general relativity but deal with a much wider class of modifications of the Hamiltonian constraint than previously. Showing that these modifications render the Hamiltonian constraint second class can be seen as evidence for the statement that the demand of closure of the constraint algebra is consequential for  Lorentz invariance. Moreover, the former  is a criterion applicable in the full, non linear setting, unlike Lorentz invariance. 
 
 It is also worth mentioning attempts to explicitly deform the hyper surface deformation algebra due to Bojowald et. al. (see \cite{Bojowald}) who considered the `deformed' Poisson bracket of the Hamiltonian constraint with itself, which reads
 \begin{equation}\left\{H(N),H(M)\right\}=-H_{a}(\beta g^{ab}(N\nabla_{b}M-M\nabla_{b}N)),\label{Bojowald}\end{equation}
 where $\beta$ is the deformation  thought to arise from inverse triad or holonomy corrections in the effective equations of motion (in loop variables). 
 The function $\beta$ is not only a function of the phase space variables but also of additional quantities which become available in an effective formulation of Quantum Gravity. However, our analysis deals only with classical phase space variables and hence has no bearing on the above scenario and its viability at the full non linear level.

In \cite{Martinec}, an arbitrary extension of the potential term was considered, but the assumption of an ultra-local kinetic term was retained, and no linear term in the momentum was allowed. We extend both of these assumptions.  

In \cite{XG}, \cite{Saitou}, and references therein, the Hamiltonian analysis of spatially covariant theories of gravity is carried out. These theories are expected to propagate a scalar degree of freedom in addition to the gravitational ones, which would have them propagate a total of three degrees of freedom in 3+1 dimensions. Our end results show that the class of modifications of the Hamiltonian constraint we consider render it second class, so it is natural to expect that the theory described by these modifications of the constraint is among those studied in the aforementioned works.

  In \cite{SD_uniqueness}, following previous work on a BRST treatment of pairs of first class constraints which were gauge-fixing each other, one of used the following criteria for the selection of two sets of constraints (one gauge-fixing the other) : i)  The constraints must be scalar. That is, they must constrain one degree of freedom per space point and that the emerging gauge-fixed theory possess two \emph{dynamically} propagating metric degrees of freedom per space point. 
This also implied that each constraint  was first class when taken in conjunction with the spatial diffeomorphism constraint. ii)
one of the constraint should be second-class wrt the other,  up to a finite-dimensional kernel. This requirement just means that each will serve as a good gauge-fixing for the other and that observables can be related to two different symmetries. 
The further important point about this requirement, is that it would be the one responsible for eliminating higher order theories (for instance Weyl gravity). This occurs because the formal manifestation of this requirement -- that the bracket be everywhere invertible -- is implemented firstly by the elliptic character of the operator $\{\xi_1, \xi_2\}$. For higher order theories (e.g. with constraints depending on higher order curvature and momentum), the operator will always be phase space dependent, and generically non-positive. 

\section{Separation of Orders of Spatial Derivatives }

As we saw in the previous section, utilising the fact that terms of
different spatial derivative order do not interfere with one another
allows us to further simplify our analysis. We can in principle write 

\[
F[g,\pi,x)=\sum_{k=2}^{q}F^{(k)}[g,\pi,x),
\]
 where each $F^{(k)}[g,\pi,x)$ contains $k$ spatial derivatives.
The lowest number of derivatives is $k=2$, because the kinetic term
is a scalar, and the two momenta have an even number of indices, which need to be contracted.  The only covariant object with a single index we can
form from the metric and its derivatives is the covariant derivative,  which, minimally, would need to be contracted with itself, in a term such as
$\pi^{ab}\nabla^{2}\pi_{ab}$ or $\nabla_{i}\pi_{ab}\nabla^{i}\pi^{ab}$.  Alternatively, the only covariant tensor containing any derivatives which one can form solely from the metric
 is the curvature tensor, containing
 two spatial derivatives. This term can feature in a modification
such as $R_{ij}\pi^{ij}\pi$ or $RG_{abcd}\pi^{ab}\pi^{cd}$. In due course we will write down the most general form of such
modifications. 

At this point, we have:
\be\label{mixed_kin_bracket1}
\sum_{k=2}^{q}\left(\{F^{(k)}(f),F_{o}(h)\}-\{F^{(k)}(h),F_{o}(f)\}\right)
\ee
 
\[
=\{F^{(2)}(f),F_{o}(h)\}-\{F^{(2)}(h),F_{o}(f)\}+\sum_{k=4}^{q}\left(\{F^{(k)}(f),F_{o}(h)\}-\{F^{(k)}(h),F_{o}(f)\}\right).
\]
We have singled out the first term of \eqref{mixed_kin_bracket1} because there is no
potential for interference between this term and any term coming from
the bracket $\{\sum_{k=2}^{q}F^{(k)}(f),\sum_{k'=2}^{q}F^{(k')}(h)\}$,
whereas a term such as $\{F^{(4)}(f),F_{o}(h)\}-\{F^{(4)}(h),F_{o}(f)\}$
is of the same order as $\{F^{(2)}(f),F^{(2)}(h)\}$, the latter coming
from the $\{F(f),F(h)\}$ bracket. In other words, although it is true that the highest order terms, represented  in table \ref{table}, don\rq{}t mix, it could still be true that lower orders of the last line of that table would mix. However, this would not happen for the term we have separated out in \eqref{mixed_kin_bracket1}.   

Suppose we prove that the only
way this singled out term can vanish on the constraint hypersurface
is if $F^{(2)}[g,\pi,x)=0$. Then, we can move on to the next order in difficulty,  $\{F^{(4)}(f),F_{o}(h)\}-\{F^{(4)}(h),F_{o}(f)\}$. 
Now it too stands alone and un-interfered with, because $\{F^{(2)}(f),F^{(2)}(h)\}$
identically vanishes. Suppose then that we can also prove that $F^{(4)}[g,\pi,x)=0$ is the only possible way for the constraint
to be first class, then we would have to consider $\{F^{(6)}(f),F_{o}(h)\}-\{F^{(6)}(h),F_{o}(f)\}$,
so on and so forth. We therefore find that  to check whether or not the 
Hamiltonian constraint second class it suffices to 
consider 

\[
\{F^{(k)}(f),F_{o}(h)\}-\{F^{(k)}(h),F_{o}(f)\}
\]
 where $0<k\leq q$, assuming $F^{(k')}[g,\pi,x)=0$ $\forall k'<k.$ 
 
 Given that we are
demanding that this bracket vanish on the constraint hypersurface,
it is worth noting that the separation in orders of spatial gradients,
as well as the way in which Poisson brackets maintain this order, is already sufficient 
to disallow the appearance of terms proportional to the Hamiltonian constraint  on the
right hand side of the expression. One can see this in the following way. First, apart
from maintaining the order of spatial derivatives, the Poisson bracket
 necessarily reduces the polynomial power of the momenta by one.
The kinetic term is quadratic in momenta and the potential is independent
of the momenta. Without calculating anything, we can already say from
this rule that $\left\{ F(f),F_{o}(h)\right\} -\left\{ F(h),F_{o}(f)\right\} ,\ \left\{ F(f),F(h)\right\} $
will result in expressions cubic in the momenta and $\left\{ F_{o}(f),\hat{V}(h)\right\} +\left\{ \hat{V}(f),F_{o}(h)\right\} ,\ \left\{ F(f),\hat{V}(h)\right\} +\left\{ \hat{V}(f),F(h)\right\} $
will result in expressions linear in the momenta. The only  hope of getting a Hamiltonian constraint on the rhs of the bracket then would
be that these expressions combine and one power of the momentum can
be factored out to form the (deformed) Hamiltonian constraint. Focusing on $\left\{ F^{(q)}(f),F_{o}(h)\right\} -\left\{ F^{(q)}(h),F_{o}(f)\right\} $,
-- from which we would be required to extract the kinetic part of the constraint, $F_{o}[g,\pi,x)+F[g,\pi,x)$ --  
we would need an expression cubic in the momenta containing the ultralocal
piece $F_{o}[g,\pi,x)$, i.e. a term of the form $F_{o}[g,\pi,x)A[g,\pi;x)$ for some functional of the metric linear in the momentum, $A$. Now, since there are $q$ spatial derivatives at most that are
present, the third momentum tensor in that cubic term will have all
$q$ derivatives acting on it. However, to complete the Hamiltonian constraint, we still need to obtain the  $F[g,\pi,x),$ part, which also needs to be of the form
$F_{o}[g,\pi,x)A[g,\pi;x)$.  But this is not possible, by virtue of the fact that all
$q$ of the available spatial derivatives were utilised in forming
$F[g,\pi,x)$ and there are none left for $A$.  Thus we see that schematically (suppressing indices),
we could at most have, for some $B[g,\pi;x), C[g,\pi;x), $ for some functional of the metric linear in the momentum and with no derivatives, $B$, and general $C$,

\[
\underset{q ~ \mbox{derivatives}}{\underbrace{A[g,\pi;x)}}F_{o}[g,\pi,x)+\underset{0 ~ \mbox{derivatives}}{\underbrace{B[g,\pi;x)}}F[g,\pi,x)\neq C[g,\pi;x)\left(F_{o}[g,\pi,x)+F[g,\pi,x)\right)
\]
In summary, due to the  disparate numbers of spatial derivatives, we cannot match $A$ and $B$,  dashing any
hopes of forming the Hamiltonian constraint. In conclusion, we can at most 
have

\begin{equation}
\left\{ F^{(k)}(x),F_{o}(y)\right\} -\left\{ F^{(k)}(x),F_{o}(y)\right\} =\int\textrm{d}^{3}z\big({}^{(k)}Y_{H(x)H(y)}^{H_{a}(z)}[g,\pi;,x,y)H^{a}(z)\big).
\end{equation}
Here $^{(k)}Y_{H(x)H(y)}^{H_{a}(z)}[g,\pi;x,y)$ is contained in $U_{H(x)H(y)}^{H_{a}(z)}[g,\pi;x,y)$, given in \eqref{general_first_class}.

\section{Computing the Bracket}

In this section, we will delve into the explicit calculation of $\left\{ F^{(k)}(f),F_{o}(h)\right\} -\left\{ F^{(k)}(h),F_{o}(f)\right\} $,
and prove that the first class nature of the Hamiltonian constraint
survives no modification of the kinetic term. Recall that

\[
\left\{ F^{(k)}(f),F_{o}(h)\right\} -\left\{ F^{(k)}(h),F_{o}(f)\right\} =\int \textrm{d}^{3}x\left(\frac{\delta F^{(k)}(f)}{\delta g_{ij}(x)}\frac{\delta F(h)}{\delta\pi^{ij}(x)}-\frac{\delta F^{(k)}(f)}{\delta\pi^{ij}(x)}\frac{\delta F(h)}{\delta g_{ij}(x)}\right)-(f\leftrightarrow h),
\]
 where, $\left(f\leftrightarrow h\right)$ means we evaluate the same
expression but after swapping the smearing functions $f$ and $h$. 
We can very easily compute the functional derivatives of $F[g,\pi,x)$
with respect to the canonical variables to find:

\[
\left(\frac{\delta F^{(k)}(f)}{\delta g_{ij}}\cdot\frac{\delta F(h)}{\delta\pi^{ij}}-\frac{\delta F^{(k)}(f)}{\delta\pi^{ij}}\cdot\frac{\delta F(h)}{\delta g_{ij}}\right)-(f\leftrightarrow h)=
\]

\be\label{equ:intermediate_PB}
\int \textrm{d}^{3}x\left(\left(\frac{h(x)}{\sqrt{g}}(\pi_{ij}-\frac{1}{2}g_{ij}\pi).\frac{\delta F^{(k)}(f)}{\delta g_{ij}(x)}\right)-\left(\frac{\delta F^{(k)}(f)}{\delta\pi^{cd}(x)}\cdot\frac{h(x)}{\sqrt{g}}(\pi^{kc}\pi_{k}^{~d}-\frac{1}{2}\pi\pi^{cd})\right)\right)-(f\leftrightarrow h).
\ee
The functional derivatives of $F^{(k)}[g,\pi,x)$ with respect to
the canonical variables are much more complicated. 

 Before we compute it,  we should note that this calculation doesn't necessarily require
the consideration of \emph{all} the terms resulting from brackets. When we compute the bracket, we will first consider the variations $\delta_{g}F^{(k)}[g,\pi,x)$ and $\delta_{\pi}F^{(k)}[g,\pi,x)$. We find that where ever in these variations that a term which contains no spatial derivatives acting on either of $\delta g_{ab}$ or $\delta \pi^{ab}$ are produced, such terms will not contribute to the Poisson bracket. This is because we will take the product of the functional derivatives that defines the Poisson brackets and then subtract from that product the same expression with the smearing functions $f(x)$ and $h(x)$ interchanged. This gives rise to terms containing expression such as $(f\nabla\cdots\nabla h-h\nabla \cdots \nabla f)$ within an integral, and the derivatives on the smearing functions are those that were on either $\delta g_{ab}$ or $\delta \pi^{ab}$ and after subsequent integration by parts were moved on to the smearing functions. Clearly, if there were no derivatives at all which were to be moved away from either of $\delta g_{ab}$, $\delta \pi^{ab}$ on to the smearing functions, the expression under the integral would be $fh-hf=0$ and thus will not contribute to the Poisson bracket at all. 

Note that in the bracket we are specifically interested in, all the spatial derivatives  originate from $F^{(k)}[g,\pi,x)$ because the other term $F_{o}[g,\pi,x)$ is ultra local in both the metric and the momenta which means it contains no spatial gradients at all. Moreover, this is the reason why $F_{o}[g,\pi,x)$ commutes with itself.

It is fairly simple to compute $\delta_{\pi}F^{(k)}[g,\pi,x)$:

\[
\delta_{\pi}F^{(k)}[g,\pi,x)=\delta_{\pi}(\mathcal{B}_{abcd}^{i_{1}\cdots i_{n}\,j_{1}\cdots j_{m}}\left(\nabla_{i_{1}}\cdots\nabla_{i_{n}}\pi^{cd}\right)\left(\nabla_{j_{1}}\cdots\nabla_{j_{m}}\pi^{ab}\right))
\]
 
\[
\mathcal{B}_{abcd}^{i_{1}\cdots i_{n}\,j_{1}\cdots j_{m}}\left(\nabla_{i_{1}}\cdots\nabla_{i_{n}}\delta\pi^{cd}\right)\left(\nabla_{j_{1}}\cdots\nabla_{j_{m}}\pi^{ab}\right)+\mathcal{B}_{abcd}^{i_{1}\cdots i_{n}\,j_{1}\cdots j_{m}}\left(\nabla_{i_{1}}\cdots\nabla_{i_{n}}\pi^{cd}\right)\left(\nabla_{j_{1}}\cdots\nabla_{j_{m}}\delta\pi^{ab}\right).
\]

 Note that $k=max(m,n,r'+2).$ The contribution from this term to the
Poisson bracket then reads:

\[
-\left(\frac{\delta F^{(k)}(f)}{\delta\pi^{cd}}\cdot\frac{h(x)}{\sqrt{g}}(\pi^{kc}\pi_{k}^{~d}-\frac{1}{2}\pi\pi^{cd})\right)-(f\leftrightarrow h)=
\]

\[
-\int\textrm{d}^{3}x\frac{1}{{g}}\nabla_{i_{1}}\cdots\nabla_{i_{n}}\left(f(x)\mathcal{B}_{abcd}^{i_{1}\cdots i_{n}\,j_{1}\cdots j_{m}}\left(\nabla_{j_{1}}\cdots\nabla_{j_{m}}\pi^{ab}\right)\right)\left(h(x)(\pi^{kc}\pi_{k}^{~d}-\frac{1}{2}\pi\pi^{cd})\right)+
\]
 
\[
+(-1)^{n}\int\textrm{d}^{3}x\frac{1}{{g}}h(x)\mathcal{B}_{abcd}^{i_{1}\cdots i_{n}\,j_{1}\cdots j_{m}}\left(\nabla_{j_{1}}\cdots\nabla_{j_{m}}\pi^{ab}\right)\nabla_{i_{n}}\cdots\nabla_{i_{1}}\left(f(x)(\pi^{kd}\pi_{k}^{~c}-\frac{1}{2}\pi\pi^{cd})\right)+
\]

\[
+\{(\nabla_{i_{1}},\cdots,\nabla_{i_{n}},cd)\leftrightarrow(\nabla_{j_{1}},\cdots,\nabla_{j_{n}},ab)\}-(f\leftrightarrow h).
\]
 Determining the metric variation will be more tedious
given the nature of the metric variation of the kinetic term. 

\subsection{Brackets involving variations of the generalized super-metric, $\mathcal{B}_{abcd}^{i_{i}\cdots i_{n}j_{1}\cdots j_{m}}$.}
First,
let us consider the metric variation of $\mathcal{B}_{abcd}^{i_{i}\cdots i_{n}j_{1}\cdots j_{m}}$.
This is a function of

\[
R_{kl},\ \nabla_{a_{1}}R_{kl},\ \nabla_{a_{1}}\nabla_{a_{2}}R_{kl},\ \nabla_{a_{1}}\nabla_{a_{2}}\cdots\nabla_{a_{r'}}R_{kl},
\]
 so we can write

\begin{eqnarray*}
\delta_{g}\mathcal{B}_{abcd}^{i_{i}\cdots i_{n}j_{1}\cdots j_{m}}=\frac{\partial\mathcal{B}_{abcd}^{i_{i}\cdots i_{n}j_{1}\cdots j_{m}}}{\partial R_{kl}}\delta_{g}(R_{kl})+\frac{\partial\mathcal{B}_{abcd}^{i_{i}\cdots i_{n}j_{1}\cdots j_{m}}}{\partial(\nabla_{a_{1}}R_{kl})}\delta_{g}(\nabla_{a_{1}}R_{kl})+\\
+\frac{\partial\mathcal{B}_{abcd}^{i_{i}\cdots i_{n}j_{1}\cdots j_{m}}}{\partial(\nabla_{a_{1}}\nabla_{a_{2}}R_{kl})}\delta_{g}(\nabla_{a_{1}}\nabla_{a_{2}}R_{kl})+\cdots+\frac{\partial\mathcal{B}_{abcd}^{i_{i}\cdots i_{n}j_{1}\cdots j_{m}}}{\partial(\nabla_{a_{1}}\nabla_{a_{2}}\cdots\nabla_{a_{r'}}R_{kl})}\delta_{g}(\nabla_{a_{1}}\nabla_{a_{2}}\cdots\nabla_{a_{r'}}R_{kl}),
\end{eqnarray*}
 then the objective is to write down each of the above variations
in terms of the variation of the metric tensor. This is the functional analogue of the chain rule which will be fruitful to utilise here. This is the same method of computing variations as that which was used in \cite{XG}. The variations of
the curvature tensor and its derivatives, which are given as follows:

\begin{eqnarray*}
\ \delta R_{kl}=^{(0)}\mathcal{\mathcal{\mathcal{A}}}_{kl}^{ijl_{1}l_{2}}\nabla_{l_{1}}\nabla_{l_{2}}\delta g_{ij},\ \ \ \ \ \ \ \ \ \ \ \ \ \ \ \ \ \ \ \ \ \ \ \ \ \ \ \ \ \ \ \ \ \ \ \ \ \\
\delta\left(\nabla_{a_{1}}R_{kl}\right)=^{(0)}\mathcal{\mathcal{\mathcal{A}}}_{kl}^{ijl_{1}l_{2}}\nabla_{a_{1}}\nabla_{l_{1}}\nabla_{l_{2}}\delta g_{ij}-^{(1)}\mathcal{A}_{a_{1}kl}^{ijl'}\nabla_{l'}\delta g_{ij},\ \ \ \ \ \ \ \ \ \ \ \ \ \ \ \ \ \ \ \ \ \ \ \ \ \ \ \ \ \ \ \ \ \ \ \ \ \\
\delta(\nabla_{a_{1}}\nabla_{a_{2}}R_{kl})=^{(0)}\mathcal{\mathcal{\mathcal{A}}}_{kl}^{ijl_{1}l_{2}}\nabla_{a_{1}}\nabla_{a_{2}}\nabla_{l_{1}}\nabla_{l_{2}}\delta g_{ij}-^{(1)}\mathcal{A}_{a_{1}kl}^{ijl'}\nabla_{a_{2}}\nabla_{l'}\delta g_{ij}-^{(2)}\mathcal{A}_{a_{1}a_{2}kl}^{ijl'}\nabla_{l'}\delta g_{ij},\ \ \ \ \ \ \ \ \ \ \ \ \ \ \ \ \ \ \ \ \ \ \ \ \ \ \ \ \ \ \ \ \ \ \ \ \ \\
\cdot\ \ \ \ \ \ \ \ \ \ \ \ \ \ \ \ \ \ \ \ \ \ \ \ \ \ \ \ \ \ \ \ \ \ \ \ \ \ \ \ \ \ \ \ \ \ \ \ \ \ \ \ \ \ \ \ \ \ \ \ \ \ \ \ \ \ \ \ \ \ \ \ \ \ \\
\cdot\ \ \ \ \ \ \ \ \ \ \ \ \ \ \ \ \ \ \ \ \ \ \ \ \ \ \ \ \ \ \ \ \ \ \ \ \ \ \ \ \ \ \ \ \ \ \ \ \ \ \ \ \ \ \ \ \ \ \ \ \ \ \ \ \ \ \ \ \ \ \ \ \ \ \\
\cdot\ \ \ \ \ \ \ \ \ \ \ \ \ \ \ \ \ \ \ \ \ \ \ \ \ \ \ \ \ \ \ \ \ \ \ \ \ \ \ \ \ \ \ \ \ \ \ \ \ \ \ \ \ \ \ \ \ \ \ \ \ \ \ \ \ \ \ \ \ \ \ \ \ \ \\
\delta(\nabla_{a_{1}}\nabla_{a_{2}}\cdots\nabla_{a_{r'}}R_{kl})=^{(0)}\mathcal{\mathcal{\mathcal{A}}}_{kl}^{ijl_{1}l_{2}}\nabla_{a_{1}}\nabla_{a_{2}}\cdots\nabla_{a_{r'}}\nabla_{l_{1}}\nabla_{l_{2}}\delta g_{ij}-^{(1)}\mathcal{A}_{a_{1}kl}^{ijl'}\nabla_{a_{2}}\nabla_{a_{3}}\cdots\nabla_{a_{r'}}\nabla_{l'}\delta g_{ij},\ \ \ \ \ \ \ \ \ \ \ \ \ \ \ \ \ \ \ \ \ \ \ \ \ \ \ \ \ \ \ \ \ \ \ \ \ \\
-^{(2)}\mathcal{A}_{a_{1}a_{2}kl}^{ijl'}\nabla_{a_{3}}\nabla_{a_{4}}\cdots\nabla_{a_{r'}}\nabla_{l'}\delta g_{ij}-\cdots-^{(n)}\mathcal{A}_{a_{1}a_{2}\cdots a_{r'}kl}^{ijl'}\nabla_{l'}\delta g_{ij}.\ \ \ \ \ \ \ \ \ \ \ \ \ \ \ \ \ \ \ \ \ \ \ \ \ \ \ \ \ \ \ \ \ \ \ \ \ 
\end{eqnarray*}
 For the structure of the tensors $^{(0)}\mathcal{A}_{kl}^{ijl_{1}l_{2}},\ {}^{(1)}\mathcal{A}_{a_{1}kl}^{ijl'},\ ^{(2)}\mathcal{A}_{a_{1}a_{2}kl}^{ijl'},\cdots,\ ^{(n)}\mathcal{A}_{a_{1}a_{2}\cdots a_{r'}kl}^{ijl'}$
to be shown, it will help to define the object 

\begin{equation}
\Xi_{nkl}^{mij}=\frac{1}{2}\left(g_{k}^{m}g_{n}^{(i}g_{k}^{j)}-g_{n}^{m}g_{k}^{(i}g_{l}^{j)}\right).
\label{Xi_uncontracted}\end{equation}
 Now, we can define the following 

\begin{equation}
^{(0)}\mathcal{A}_{kl}^{ijl_{1}l_{2}}=g^{l_{1}k'}\Xi_{k'kl}^{l_{2}ij}-g^{k_{1}k_{2}}\Xi_{k_{1}k_{2}(k}^{l_{2}ij}g_{l)}^{l_{1}},
\end{equation}

\begin{equation}
^{(1)}\mathcal{A}_{a_{1}kl}^{ijl'}=R_{k}^{k'}\Xi_{k'la_{1}}^{l'ij}+R_{l}^{k'}\Xi_{k'ka_{1}}^{l'ij},
\end{equation}

\begin{equation}
^{(2)}\mathcal{A}_{a_{1}a_{2}kl}^{ijl'}=\nabla^{k'}R_{kl}\Xi_{k'a_{1}a_{2}}^{l'ij}+\nabla_{a_{1}}R_{k}^{k'}\Xi_{k'la_{2}}^{l'ij}+\nabla_{a_{1}}R_{l}^{k'}\Xi_{k'ka_{2}}^{l'ij}+\nabla_{a_{2}}R_{k}^{k'}\Xi_{k'la_{1}}^{l'ij}+\nabla_{a_{2}}R_{l}^{k'}\Xi_{k'ka_{1}}^{l'ij}.
\end{equation}
The rest of the tensors $^{(k)}\mathcal{A}_{a_{1}\cdots a_{k}kl}^{ijl'}$ for $2<k\leq n$ are also functions of the tensor $\Xi$ and $\nabla_{a_{1}}\cdots \nabla_{a_{k-2}}R_{kl}$. Their form will not be necessary for the purposes of our proof and so we leave won't write down the explicit form of these functions.
 
 Now, the contribution from these variations to the Poisson bracket
is:

\[
\left(\frac{h(x)}{\sqrt{g}}(\pi_{ij}-\frac{1}{2}g_{ij}\pi).\frac{\delta F^{(k)}(f)}{\delta g_{ij}}\right)-(f\leftrightarrow h)\supset
\]

\[\int\textrm{d}^{3}x
\bigg[\bigg\{\nabla_{l_{1}}\nabla_{l_{2}}\left(\frac{f}{\sqrt{g}}\frac{\partial\left(\mathcal{B}_{abcd}^{i_{1}\cdots i_{n}\,j_{1}\cdots j_{m}}\right)}{\partial R_{kl}}\left(\nabla_{i_{1}}\cdots\nabla_{i_{n}}\pi^{cd}\right)\left(\nabla_{j_{1}}\cdots\nabla_{j_{m}}\pi^{ab}\right){}^{(0)}\mathcal{A}_{kl}^{ijl_{1}l_{2}}\right)+
\]

\[
+\nabla_{l'}\left(\frac{f}{\sqrt{g}}\frac{\partial\left(\mathcal{B}_{abcd}^{i_{1}\cdots i_{n}\,j_{1}\cdots j_{m}}\right)}{\partial(\nabla_{a_{1}}R_{kl})}\left(\nabla_{i_{1}}\cdots\nabla_{i_{n}}\pi^{cd}\right)\left(\nabla_{j_{1}}\cdots\nabla_{j_{m}}\pi^{ab}\right){}^{(1)}\mathcal{A}_{a_{1}kl}^{ijl'}\right)-
\]

\[
-\nabla_{l_{2}}\nabla_{l_{1}}\nabla_{a_{1}}\left(\frac{f}{\sqrt{g}}\frac{\partial\left(\mathcal{B}_{abcd}^{i_{1}\cdots i_{n}\,j_{1}\cdots j_{m}}\right)}{\partial(\nabla_{a_{1}}R_{kl})}\left(\nabla_{i_{1}}\cdots\nabla_{i_{n}}\pi^{cd}\right)\left(\nabla_{j_{1}}\cdots\nabla_{j_{m}}\pi^{ab}\right){}^{(0)}\mathcal{A}_{kl}^{ijl_{1}l_{2}}\right)+
\]

\[
+\nabla_{l'}\left(\frac{f}{\sqrt{g}}\frac{\partial\left(\mathcal{B}_{abcd}^{i_{1}\cdots i_{n}\,j_{1}\cdots j_{m}}\right)}{\partial(\nabla_{a_{1}}\nabla_{a_{2}}R_{kl})}\left(\nabla_{i_{1}}\cdots\nabla_{i_{n}}\pi^{cd}\right)\left(\nabla_{j_{1}}\cdots\nabla_{j_{m}}\pi^{ab}\right){}^{(2)}\mathcal{A}_{a_{1}a_{2}kl}^{ijl'}\right)-
\]

\[
-\nabla_{l'}\nabla_{a_{1}}\left(\frac{f}{\sqrt{g}}\frac{\partial\left(\mathcal{B}_{abcd}^{i_{1}\cdots i_{n}\,j_{1}\cdots j_{m}}\right)}{\partial(\nabla_{a_{1}}\nabla_{a_{2}}R_{kl})}\left(\nabla_{i_{1}}\cdots\nabla_{i_{n}}\pi^{cd}\right)\left(\nabla_{j_{1}}\cdots\nabla_{j_{m}}\pi^{ab}\right){}^{(1)}\mathcal{A}_{a_{2}kl}^{ijl'}\right)
\]

\[
+\nabla_{l_{2}}\nabla_{l_{1}}\nabla_{a_{2}}\nabla_{a_{1}}\left(\frac{f}{\sqrt{g}}\frac{\partial\left(\mathcal{B}_{abcd}^{i_{1}\cdots i_{n}\,j_{1}\cdots j_{m}}\right)}{\partial(\nabla_{a_{1}}\nabla_{a_{2}}R_{kl})}\left(\nabla_{i_{1}}\cdots\nabla_{i_{n}}\pi^{cd}\right)\left(\nabla_{j_{1}}\cdots\nabla_{j_{m}}\pi^{ab}\right){}^{(0)}\mathcal{A}_{kl}^{ijl_{1}l_{2}}\right)+\cdots+
\]

\[
+\epsilon\nabla_{l_{2}}\nabla_{l_{1}}\nabla_{a_{r'}}\nabla_{a_{r-1}}\cdots\nabla_{a_{1}}\left(\frac{f}{\sqrt{g}}\frac{\partial\left(\mathcal{B}_{abcd}^{i_{1}\cdots i_{n}\,j_{1}\cdots j_{m}}\right)}{\partial(\nabla_{a_{1}}\nabla_{a_{2}}\cdots\nabla_{a_{r'}}R_{kl})}\left(\nabla_{i_{1}}\cdots\nabla_{i_{n}}\pi^{cd}\right)\left(\nabla_{j_{1}}\cdots\nabla_{j_{m}}\pi^{ab}\right){}^{(0)}\mathcal{A}_{kl}^{ijl_{1}l_{2}}\right)+\cdots+
\]

\[
+\epsilon'\nabla_{l'}\left(\frac{f}{\sqrt{g}}\frac{\partial\left(\mathcal{B}_{abcd}^{i_{1}\cdots i_{n}\,j_{1}\cdots j_{m}}\right)}{\partial(\nabla_{a_{1}}\nabla_{a_{2}}\cdots\nabla_{a_{r'}}R_{kl})}\left(\nabla_{i_{1}}\cdots\nabla_{i_{n}}\pi^{cd}\right)\left(\nabla_{j_{1}}\cdots\nabla_{j_{m}}\pi^{ab}\right){}^{(n)}\mathcal{A}_{a_{1}a_{2}\cdots a_{r'}kl}^{ijl'}\right)\bigg\}\times
\]

\begin{equation}
\times h\left(\pi_{ij}-\frac{1}{2}g_{ij}\pi\right)\bigg]-(f\leftrightarrow h).
\end{equation}
 This expression requires us to integrate by parts until all derivatives
hit the derivative of the function $\mathcal{B}_{abcd}^{i_{1}\cdots i_{n}\,j_{1}\cdots j_{m}}$
with respect to the derivatives of the Ricci tensor multiplied by
the smearing function. This is why the sign in front of each monomial
varies depending on the order of derivatives of the Ricci tensor we
are considering, and since we keep the highest order ($r'+2$) arbitrary,
we do not know if it is even or odd and hence we have multiplied the
last couple of monomials by factors $\epsilon$ and $\epsilon'$.
Our main interest, for reasons we will later explain, will be with
the terms containing the highest number of covariant derivatives.
For the purposes
of the proof, we will only be using the terms containing the
highest order of covariant derivatives acting on the momenta or on the smearing functions. 

This means that out of this calculation,
we will pay heed only to the term: 

$$
\int \textrm{d}^{3}x\bigg[\epsilon\nabla_{l_{2}}\nabla_{l_{1}}\nabla_{a_{r'}}\nabla_{a_{r-1}}\cdots\nabla_{a_{1}}\left(\frac{f}{\sqrt{g}}\frac{\partial\left(\mathcal{B}_{abcd}^{i_{1}\cdots i_{n}\,j_{1}\cdots j_{m}}\right)}{\partial(\nabla_{a_{1}}\nabla_{a_{2}}\cdots\nabla_{a_{r'}}R_{kl})}\left(\nabla_{i_{1}}\cdots\nabla_{i_{n}}\pi^{cd}\right)\left(\nabla_{j_{1}}\cdots\nabla_{j_{m}}\pi^{ab}\right){}^{(0)}\mathcal{A}_{kl}^{ijl_{1}l_{2}}\right)\times
$$
\be
\times h\left(\pi_{ij}-\frac{1}{2}g_{ij}\pi\right)\bigg]-(f\leftrightarrow h).
\label{rterm}
\ee
Also for the purposes of the proof, we will choose to localise this above integral by setting one of the smearing functions $h(x)$ to a Dirac delta function, leading to the expression:
$$
\nabla_{l_{2}}\nabla_{l_{1}}\nabla_{a_{r'}}\cdots\nabla_{a_{1}}\left({f}\frac{\partial\left(\mathcal{B}^{i_1\cdots i_n\,j_1\cdots j_m}_{abcd}\right)}{\partial(\nabla_{a_{1}}\nabla_{a_{2}}\cdots\nabla_{a_{r'}}R_{kl})}\nabla_{i_1}\cdots \nabla_{i_n}\pi^{cd}\nabla_{j_1}\cdots \nabla_{j_m}\pi^{ab}\right)\left(\,^{(0)}\mathcal{A}_{kl}^{ijl_{1}l_{2}}G_{ijef}\pi^{ef}\right)$$ 
$$
-(-1)^{r'}\left(\frac{\partial\left(\mathcal{B}^{i_1\cdots i_n\,j_1\cdots j_m}_{abcd}\right)}{\partial(\nabla_{a_{1}}\nabla_{a_{2}}\cdots\nabla_{a_{r'}}R_{kl})}\nabla_{i_1}\cdots \nabla_{i_n}\pi^{cd}\nabla_{j_1}\cdots \nabla_{j_m}\pi^{ab}\right) \nabla_{a_1}\cdots\nabla_{a_{r'}}\nabla_{l_{1}}\nabla_{l_{2}}\left(f(x)\,^{(0)}\mathcal{A}_{kl}^{ijl_{1}l_{2}}G_{ijef}\pi^{ef}\right).
$$
It will also be useful to collect all terms containing $\left(\,^{(0)}\mathcal{A}_{kl}^{ijl_{1}l_{2}}G_{ijef}\pi^{ef}\right)$. There will be one such term at every order of derivatives and symbolically this sum takes a simple form:
$$\sum^{r'}_{r=0}\bigg[
\nabla_{l_{2}}\nabla_{l_{1}}\nabla_{a_{r}}\cdots\nabla_{a_{1}}\left({f}\frac{\partial\left(\mathcal{B}^{i_1\cdots i_n\,j_1\cdots j_m}_{abcd}\right)}{\partial(\nabla_{a_{1}}\nabla_{a_{2}}\cdots\nabla_{a_{r}}R_{kl})}\nabla_{i_1}\cdots \nabla_{i_n}\pi^{cd}\nabla_{j_1}\cdots \nabla_{j_m}\pi^{ab}\right)\left(\,^{(0)}\mathcal{A}_{kl}^{ijl_{1}l_{2}}G_{ijef}\pi^{ef}\right)$$ 
$$
-(-1)^{r}\left(\frac{\partial\left(\mathcal{B}^{i_1\cdots i_n\,j_1\cdots j_m}_{abcd}\right)}{\partial(\nabla_{a_{1}}\nabla_{a_{2}}\cdots\nabla_{a_{r}}R_{kl})}\nabla_{i_1}\cdots \nabla_{i_n}\pi^{cd}\nabla_{j_1}\cdots \nabla_{j_m}\pi^{ab}\right) \nabla_{a_1}\cdots\nabla_{a_{r}}\nabla_{l_{1}}\nabla_{l_{2}}\left(f(x)\,^{(0)}\mathcal{A}_{kl}^{ijl_{1}l_{2}}G_{ijef}\pi^{ef}\right)\bigg].
$$
The reason for bundling up these terms as we have will be clear in the proof that follows. 

\subsection{Brackets involving the explicit covariant derivatives acting on the momenta. }
 We will now study the full structure of the following variation:

\[
\delta_{g}(\nabla_{i_{1}}\cdots\nabla_{i_{m}}\pi^{ab}).
\]
 This term depends on the metric solely through the covariant derivatives
$\nabla_{i_{k}}$, for when this operator acts on any tensor, say
$F_{b}^{a},$ it does so as follows:

\[
\nabla_{i}F_{b}^{a}=\partial_{i}F_{b}^{a}+\Gamma_{id}^{a}F^{d}-\Gamma_{ib}^{d}F_{d}^{a},
\]
 where the Christoffel symbols $\Gamma_{bc}^{a}$ carry the explicit
metric dependence. The key variational identity that will be used
repeatedly is

\[
\delta_{g}\Gamma_{ab}^{e}=\frac{1}{2}g^{ec}(\delta_{a}^{l}\delta_{b}^{(i}\delta_{c}^{j)}+\delta_{b}^{l}\delta_{a}^{(i}\delta_{c}^{j)}-\delta_{c}^{l}\delta_{a}^{(i}\delta_{b}^{j)})\nabla_{l}\delta g_{ij}\equiv\frac{1}{2}g^{ec}\Xi_{cab}^{lij}\nabla_{l}\delta g_{ij}.
\]
 Using this, the above variation reads:

\[
\delta_{g}(\nabla_{i_{1}}\cdots\nabla_{i_{m}}\pi^{ab})=
\]
 
\[
-\delta_{g}\Gamma_{i_{1}i_{2}}^{d}\nabla_{d}\nabla_{i_{3}}\cdots\nabla_{i_{m}}\pi^{ab}-\delta_{g}\Gamma_{i_{1}i_{3}}^{d}\nabla_{i_{2}}\nabla_{d}\cdots\nabla_{i_{m}}\pi^{ab}-\delta_{g}\Gamma_{i_{1}i_{4}}^{d}\nabla_{i_{2}}\nabla_{i_{3}}\nabla_{d}\cdots\nabla_{i_{m}}\pi^{ab}-
\]

\[
\cdots-\delta_{g}\Gamma_{i_{1}i_{m}}^{d}\nabla_{i_{2}}\cdots\nabla_{d}\pi^{ab}+\delta_{g}\Gamma_{i_{1}d}^{a}\nabla_{i_{2}}\cdots\nabla_{i_{m}}\pi^{db}+\delta_{g}\Gamma_{i_{1}d}^{b}\nabla_{i_{2}}\cdots\nabla_{i_{m}}\pi^{da}-
\]
 
\[
-\nabla_{i_{1}}\delta_{g}\Gamma_{i_{2}i_{3}}^{d}\nabla_{d}\nabla_{i_{4}}\cdots\nabla_{i_{m}}\pi^{ab}-\nabla_{i_{1}}\delta_{g}\Gamma_{i_{2}i_{4}}^{d}\nabla_{i_{3}}\nabla_{d}\cdots\nabla_{i_{m}}\pi^{ab}-\cdots
\]

\[
+\nabla_{i_{1}}\delta_{g}\Gamma_{i_{2}d}^{a}\nabla_{i_{3}}\cdots\nabla_{i_{m}}\pi^{db}+\nabla_{i_{1}}\delta_{g}\Gamma_{i_{2}d}^{b}\nabla_{i_{3}}\cdots\nabla_{i_{m}}\pi^{ad}-
\]

\[
-\nabla_{i_{1}}\cdots\nabla_{i_{k}}\delta_{g}\Gamma_{i_{k+1}i_{k+2}}^{d}\nabla_{d}\nabla_{i_{k+3}}\cdots\nabla_{i_{m}}\pi^{ab}-\nabla_{i_{1}}\cdots\nabla_{i_{k}}\delta_{g}\Gamma_{i_{k+1}i_{k+3}}^{d}\nabla_{i_{k+2}}\nabla_{d}\cdots\nabla_{i_{m}}\pi^{ab}-\cdots
\]
 
\[
-\nabla_{i_{1}}\cdots\nabla_{i_{k}}\delta_{g}\Gamma_{i_{k+1}i_{m}}^{d}\nabla_{i_{k+2}}\cdots\nabla_{d}\pi^{ab}+\nabla_{i_{1}}\cdots\nabla_{i_{k}}\delta_{g}\Gamma_{i_{k+1}d}^{a}\nabla_{i_{k+2}}\nabla_{d}\cdots\nabla_{i_{m}}\pi^{db}+
\]

\[
+\nabla_{i_{1}}\cdots\nabla_{i_{k}}\delta_{g}\Gamma_{i_{k+1}d}^{b}\nabla_{i_{k+2}}\nabla_{d}\cdots\nabla_{i_{m}}\pi^{da}+\cdots
\]

\[
-\nabla_{i_{1}}\cdots\nabla_{i_{m-2}}\delta_{g}\Gamma_{i_{m-1}i_{m}}^{d}\nabla_{d}\pi^{ab}+\nabla_{i_{1}}\cdots\nabla_{i_{m-2}}\delta_{g}\Gamma_{i_{m-1}d}^{a}\nabla_{i_{m}}\pi^{db}+\nabla_{i_{1}}\cdots\nabla_{i_{m-2}}\delta_{g}\Gamma_{i_{m-1}d}^{b}\nabla_{i_{m}}\pi^{da}+
\]
 
\be\label{Christoffel_variations_all}
+\nabla_{i_{1}}\cdots\nabla_{i_{m-1}}\delta_{g}\Gamma_{i_{m}d}^{a}\pi^{db}+\nabla_{i_{1}}\cdots\nabla_{i_{m-1}}\delta_{g}\Gamma_{i_{m}d}^{b}\pi^{da}.
\ee
 The terms that have been included on the right hand side of this
expression hope to convey the general structure involving two kinds
of terms mainly. 

Our interest lies mainly in terms containing the
highest number of spatial gradients explicitly. This means that if
we were to commute derivatives into some specific form, the price
to pay will all consist of terms containing lower powers of $\nabla$
appearing explicitly. Henceforth it should be understood that derivatives
are treated as though they can be commuted freely.
In the above expression, we see that there are terms where the index of the Christoffel symbols being varied are contracted with those of the covariant derivatives to their right which we will call terms of type $\textrm{I}$. Taking the Christoffel symbol coming from the $k$th covariant derivative in $\nabla_{\left\{i\right\}}\equiv \nabla_{i_{1}}\cdots\nabla_{i_{m}}\pi^{ab}$ set, we will study their structure:
$$\delta(\nabla_{\left\{i\right\}}\pi^{ab})\supset \nabla_{i_{1}}\cdots \nabla_{i_{k-1}}\sum^{m-k}_{l=1}\delta \Gamma^{d}_{i_{k}i_{k+l}}\nabla_{i_{k+1}}\cdots \hat{\nabla}_{i_{k+l}}\nabla_{d}\cdots\nabla_{i_{m}}\pi^{ab},$$
writing this in terms of the variation of the metric, we find:
\be\label{intermediate_covds_I}\nabla_{i_{1}}\cdots \nabla_{i_{k-1}}\sum^{m-k}_{l=1}\delta \Gamma^{d}_{i_{k}i_{k+l}}\nabla_{i_{k+1}}\cdots \hat{\nabla}_{i_{k+l}}\nabla_{d}\cdots\nabla_{i_{m}}\pi^{ab}=$$ $$\nabla_{i_{1}}\cdots \nabla_{i_{k-1}}\sum^{m-k}_{l=1}g^{cd}\Xi^{pij}_{ci_{k}i_{k+l}}(\nabla_{p}\delta g_{ij})\nabla_{i_{k+1}}\cdots \hat{\nabla}_{i_{k+l}}\nabla_{d}\cdots\nabla_{i_{m}}\pi^{ab}\ee

The contribution of this term to the Poisson brackets can be found by looking at the contribution of this term to $\delta_{g}F(f)$:

$$\int \textrm{d}^{3}x f(x)\mathcal{B}^{\left\{i\right\}\left\{j\right\}}_{abrs}\nabla_{\left\{j\right\}}\pi^{rs}\bigg(\nabla_{i_{1}}\cdots \nabla_{i_{k-1}}\sum^{m-k}_{l=1}g^{cd}\Xi^{pij}_{ci_{k}i_{k+l}}(\nabla_{p}\delta g_{ij})\nabla_{i_{k+1}}\cdots \hat{\nabla}_{i_{k+l}}\nabla_{d}\cdots\nabla_{i_{m}}\pi^{ab}\bigg)$$
$$=(-1)^{k-1}\int \textrm{d}^{3}x \nabla_{i_{1}}\cdots\nabla_{i_{k-1}}\left( f(x)\mathcal{B}^{\left\{i\right\}\left\{j\right\}}_{abrs}\nabla_{\left\{j\right\}}\pi^{rs}\right) \sum^{m-k}_{l=1}g^{cd}\Xi^{pij}_{ci_{k}i_{k+l}}(\nabla_{p}\delta g_{ij})\nabla_{i_{k+1}}\cdots \hat{\nabla}_{i_{k+l}}\nabla_{d}\cdots\nabla_{i_{m}}\pi^{ab}$$

To find out how this can contribute to the bracket, we first need to take all the derivatives off the $\delta g_{ij}$ term. So we have one more integration by parts to do:
$$(-1)^{k-1}\int \textrm{d}^{3}x \nabla_{i_{1}}\cdots\nabla_{i_{k-1}}\left( f(x)\mathcal{B}^{\left\{i\right\}\left\{j\right\}}_{abrs}\nabla_{\left\{j\right\}}\pi^{rs}\right) \sum^{m-k}_{l=1}g^{cd}\Xi^{pij}_{ci_{k}i_{k+l}}(\nabla_{p}\delta g_{ij})\nabla_{i_{k+1}}\cdots \hat{\nabla}_{i_{k+l}}\nabla_{d}\cdots\nabla_{i_{m}}\pi^{ab}$$
$$=(-1)^{k}\int \textrm{d}^{3}x \nabla_{p}\bigg[\nabla_{i_{1}}\cdots\nabla_{i_{k-1}}\left( f(x)\mathcal{B}^{\left\{i\right\}\left\{j\right\}}_{abrs}\nabla_{\left\{j\right\}}\pi^{rs}\right) \sum^{m-k}_{l=1}g^{cd}\Xi^{pij}_{ci_{k}i_{k+l}}\nabla_{i_{k+1}}\cdots \hat{\nabla}_{i_{k+l}}\nabla_{d}\cdots\nabla_{i_{m}}\pi^{ab}\bigg]\delta g_{ij}.$$
If this term is to be seen as a field space one form $\int(\cdots)\delta g_{ij}$ then the Poisson bracket $\left\{\cdot,F_{o}(h)\right\}$ is to be seen as the contraction of this one form with the Hamiltonian vector field $\int (\pi_{ij}-\frac12\pi g_{ij})\frac{\delta}{\delta g_{ij}}$ to give us the contribution of this term to $\left\{F(h),F_{o}(h)\right\}$:
$$(-1)^{k}\int \textrm{d}^{3}x \nabla_{p}\bigg[\nabla_{i_{1}}\cdots\nabla_{i_{k-1}}\left( f(x)\mathcal{B}^{\left\{i\right\}\left\{j\right\}}_{abrs}\nabla_{\left\{j\right\}}\pi^{rs}\right) \sum^{m-k}_{l=1}g^{cd}\Xi^{pij}_{ci_{k}i_{k+l}}\nabla_{i_{k+1}}\cdots \hat{\nabla}_{i_{k+l}}\nabla_{d}\cdots\nabla_{i_{m}}\pi^{ab}\bigg]\times$$
\begin{equation}\times h(x)\left(\pi_{ij}-\frac{1}{2}\pi g_{ij}\right).\label{adjecent-nablas}\end{equation}
Finally we can set $h(x)=\delta(x)$ to localise the integral
\begin{equation}(-1)^{k} \nabla_{p}\bigg[\nabla_{i_{1}}\cdots\nabla_{i_{k-1}}\left( f(x)\mathcal{B}^{\left\{i\right\}\left\{j\right\}}_{abrs}\nabla_{\left\{j\right\}}\pi^{rs}\right) \sum^{m-k}_{l=1}g^{cd}\Xi^{pij}_{ci_{k}i_{k+l}}\nabla_{i_{k+1}}\cdots \hat{\nabla}_{i_{k+l}}\nabla_{d}\cdots\nabla_{i_{m}}\pi^{ab}\bigg]\left(\pi_{ij}-\frac{1}{2}\pi g_{ij}\right)\label{intermediate_Christ_I}
\end{equation}
We see here that for any intermediary  $k$, we can have at most $k=m-1$, this term can never quite yield all $m$ derivatives hitting either the smearing function $f(x)$ nor any of the momenta. But the bracket itself also requires us to compute the expression \eqref{adjecent-nablas} when we swap $f(x)$ and $h(x)$:
$$(-1)^{k}\int \textrm{d}^{3}x \nabla_{p}\bigg[\nabla_{i_{1}}\cdots\nabla_{i_{k-1}}\left( h(x)\mathcal{B}^{\left\{i\right\}\left\{j\right\}}_{abrs}\nabla_{\left\{j\right\}}\pi^{rs}\right) \sum^{m-k}_{l=1}g^{cd}\Xi^{pij}_{ci_{k}i_{k+l}}\nabla_{i_{k+1}}\cdots \hat{\nabla}_{i_{k+l}}\nabla_{d}\cdots\nabla_{i_{m}}\pi^{ab}\bigg]\times$$
$$\times f(x)\left(\pi_{ij}-\frac{1}{2}\pi g_{ij}\right)$$
$$=(-1)^{k+1}\int \textrm{d}^{3}x\bigg(\nabla_{i_{1}}\cdots\nabla_{i_{k-1}}\left( h(x)\mathcal{B}^{\left\{i\right\}\left\{j\right\}}_{abrs}\nabla_{\left\{j\right\}}\pi^{rs}\right) \sum^{m-k}_{l=1}g^{cd}\Xi^{pij}_{ci_{k}i_{k+l}}\nabla_{i_{k+1}}\cdots \hat{\nabla}_{i_{k+l}}\nabla_{d}\cdots\nabla_{i_{m}}\pi^{ab}\bigg)\times$$
$$\times \nabla_{p}\left(f(x)\left(\pi_{ij}-\frac{1}{2}\pi g_{ij}\right)\right).$$

We now integrate this by parts $k-1$ times to take all derivatives off of $h(x)$:

$$\int \textrm{d}^{3}x\left( \sum^{m-k}_{l=1} h(x)\mathcal{B}^{\left\{i\right\}\left\{j\right\}}_{abrs}\nabla_{\left\{j\right\}}\pi^{rs}\right)\times$$
$$\nabla_{i_{1}}\cdots\nabla_{i_{k-1}} \bigg[g^{cd}\Xi^{pij}_{ci_{k}i_{k+l}}\left(\nabla_{i_{k+1}}\cdots \hat{\nabla}_{i_{k+l}}\nabla_{d}\cdots\nabla_{i_{m}}\pi^{ab}\right) \nabla_{p}\left(f(x)\left(\pi_{ij}-\frac{1}{2}\pi g_{ij}\right)\right)\bigg].$$
 Now when we localise the integral by choosing $h(x)=\delta(x)$, we get:
 
 $$\left( \sum^{m-k}_{l=1}\mathcal{B}^{\left\{i\right\}\left\{j\right\}}_{abrs}\nabla_{\left\{j\right\}}\pi^{rs}\right)\times$$
\be\label{intermediate_Christ_I_swap}\nabla_{i_{1}}\cdots\nabla_{i_{k-1}} \bigg[g^{cd}\Xi^{pij}_{ci_{k}i_{k+l}}\left(\nabla_{i_{k+1}}\cdots \hat{\nabla}_{i_{k+l}}\nabla_{d}\cdots\nabla_{i_{m}}\pi^{ab}\right) \nabla_{p}\left(f(x)\left(\pi_{ij}-\frac{1}{2}\pi g_{ij}\right)\right)\bigg].\ee
As we see here, despite $k$ being arbitrary, the second line above has at most $m-1$ derivatives acting on either $\pi^{ab}$ (or $\pi_{ij}$). There still isn't either a term with $m$ derivatives of $\pi^{ab}$, since we always have a $\nabla_p$ being spent on the other momentum  (both the $k$ and $k+l$ derivative terms are missing, being offset only by the $d$ term),  nor can there be all $m$ derivatives acting on $f(x)$ (or $\pi_{ij}$) from this exchange contribution, since for mid covariant derivatives,  $k_{max}=m-1$, and thus $\order{\nabla}_{max}=k_{max}+1=m-1$. Given our method of proof, these features -- the sub-leading order in derivatives of momenta and smearing function -- have important consequences. Namely, they render terms of type I irrelevant. Moreover,  the same analysis yields sub-leading derivative order for terms of type II, \emph{with the exception of Christoffel symbols adjacent to $\pi^{ab}$}. Thus  consideration of the variations of all other Christoffel symbols become irrelevant for the proof, as we now show.  

For the terms of type $\textrm{II}$ which too come from looking at the terms in $\delta_{g}\nabla_{\left\{i\right\}}\pi^{ab}$ where the index of the Christoffel symbol of the $k$th derivative contracts with those of the momentum tensor, we have
\be \delta_{g}\nabla_{\left\{i\right\}}\pi^{ab}\supset \nabla_{i_{1}}\cdots \nabla_{i_{k-1}}\delta_{g}\Gamma^{(a}_{i_{k}d}\nabla_{i_{k+1}}\cdots \nabla_{i_{m}}\pi^{b)d}.\label{varpi} \ee
Note that there is only one term coming form the variation of the $k$th covariant derivative here unlike the sum of terms in type $\textrm{I}$. Notice also that unlike the type $\textrm{I}$ terms from \eqref{intermediate_covds_I} the index $k$ here can reach a maximum of $m$.  For other intermediary terms (i.e. $k<m$), aside from the structure of  index contractions and summations, the overall derivative ordering of \eqref{varpi} is the same as that of  \eqref{intermediate_covds_I}.  Thus for intermediate terms we will have the same sub-leading behavior as  \eqref{intermediate_Christ_I} and \eqref{intermediate_Christ_I_swap} for terms of type II. 

  So we will concentrate on that term with the hope that it yields the contribution to the Poisson bracket of the kind we are interested in. First, the contribution of the above for $k=m$ to $\delta_{g}F(f)$ is:
$$\int \textrm{d}^{3}x f(x)\mathcal{B}^{\left\{i\right\}\left\{j\right\}}_{abrs}\nabla_{\left\{j\right\}}\pi^{rs}\bigg(\nabla_{i_{1}}\cdots \nabla_{i_{m-1}}\Xi^{pij}_{ci_{m}d}(\nabla_{p}\delta g_{ij})g^{c(a}\pi^{b)d}\bigg)$$
$$=(-1)^{m}\int \textrm{d}^{3}x \nabla_{p}\bigg[\left(\nabla_{i_{1}}\cdots\nabla_{i_{m-1}}(f(x)\mathcal{B}_{abrs}^{\left\{i\right\}\left\{j\right\}}\nabla_{\left\{j\right\}}\pi^{rs})\right)\Xi^{pij}_{ci_{m}d}g^{c(a}\pi^{b)d}\bigg]\delta g_{ij}$$

We can now contract the above field space one  with $\left\{\cdot,F_{o}(h)\right\}$ to get:
$$\left\{F(f),F_{o}(h)\right\}\supset$$
\be(-1)^{m}\int \textrm{d}^{3}x \nabla_{p}\bigg[\left(\nabla_{i_{1}}\cdots\nabla_{i_{m-1}}(f(x)\mathcal{B}_{abrs}^{\left\{i\right\}\left\{j\right\}}\nabla_{\left\{j\right\}}\pi^{rs})\right)\Xi^{pij}_{ci_{m}d}g^{c(a}\pi^{b)d}\bigg]h(x)\left(\pi_{ij}-\frac{1}{2}g_{ij}\pi\right),\label{varnpi}\ee
and finally, after we localise the integral by choosing $h(x)=\delta(x)$, we get
$$\left\{F(f),F_{o}(h)\right\}\vert_{h=\delta(x)}\supset$$
$$(-1)^{m} \bigg\{\bigg(\left(\nabla_{p}\nabla_{i_{1}}\cdots\nabla_{i_{m-1}}(f(x)\mathcal{B}_{abrs}^{\left\{i\right\}\left\{j\right\}}\nabla_{\left\{j\right\}}\pi^{rs})\right)\Xi^{pij}_{ci_{m}d}g^{c(a}\pi^{b)d}\bigg)\left(\pi_{ij}-\frac{1}{2}g_{ij}\pi\right)+$$ $$+\bigg(\left(\nabla_{i_{1}}\cdots\nabla_{i_{m-1}}(f(x)\mathcal{B}_{abrs}^{\left\{i\right\}\left\{j\right\}}\nabla_{\left\{j\right\}}\pi^{rs})\right)\Xi^{pij}_{ci_{m}d}\nabla_{p}g^{c(a}\pi^{b)d}\bigg)\left(\pi_{ij}-\frac{1}{2}g_{ij}\pi\right)\bigg\}$$

Finally, a term with $m$ derivatives acting on one of the momenta and the smearing function appears, which is the one in the first of the above lines, which we shall isolate and use in further analysis. Notice that this term also contains two powers of undifferentiated momenta. It can be rewritten as:
\begin{equation}(-1)^{m} \bigg\{\left(\nabla_{i_{1}}\cdots\nabla_{i_{m}}(f(x)\mathcal{B}_{abrs}^{i_{1}\cdots \hat{i}_{m}p j_{1}\cdots j_{n}}\nabla_{j_{1}\cdots j_{n}}\pi^{rs})\right)\left(\Xi^{i_{m}ij}_{cpd}g^{c(a}\pi^{b)d}\left(\pi_{ij}-\frac{1}{2}g_{ij}\pi\right)\right)\bigg\}\label{varnablapi}\end{equation}

Now, if we go back to \eqref{varnpi}, and rewrite it with $f(x)$ swapped with $h(x)$, we find that it can be written as
\be\int \textrm{d}^{3}x \, h(x)\mathcal{B}_{abrs}^{\left\{i\right\}\left\{j\right\}}\nabla_{\left\{j\right\}}\pi^{rs}\left(\nabla_{i_{1}}\cdots\nabla_{i_{m-1}}\left[\Xi^{pij}_{ci_{m}d}g^{c(a}\pi^{b)d}\nabla_{p}\bigg[f(x)\left(\pi_{ij}-\frac{1}{2}g_{ij}\pi\right)\bigg]\right]\right)\ee

We are interested in a term containing $m$ derivatives of both the smearing function \emph{and} a linear  in momentum term with no derivatives,  when we set $h(x)=\delta(x)$ to arrive at:
\be\mathcal{B}^{i_{1}\cdots \hat{i}_{m}p j_{1}\cdots j_{n}}_{abrs}
\nabla_{j_{1}\cdots j_{n}}\pi^{rs}\, \Xi^{i_{m}ij}_{cpd}g^{c(a}\pi^{b)d}\, \nabla_{i_{1}}\cdots\nabla_{i_{m}}\left(f(x)\left(\pi_{ij}-\frac{1}{2}g_{ij}\pi\right)\right)
 \label{varnablapi2}\ee

Putting \eqref{varnablapi} and \eqref{varnablapi2} together, we find that
\[
\left\{ \left(\frac{h(x)}{\sqrt{g}}(\pi_{ij}-\frac{1}{2}g_{ij}\pi).\frac{\delta F^{(k)}(f)}{\delta g_{ij}}\right)-(f\leftrightarrow h)\right\} \vert_{\nabla^{(max)},(m>n);h(x)=\delta(x)}=
\]
$$(-1)^{m} \left(\nabla_{i_{1}}\cdots\nabla_{i_{m}}\left(f(x)\mathcal{B}_{abrs}^{i_{1}\cdots \hat{i}_{m}p j_{1}\cdots j_{n}}\nabla_{j_{1}\cdots j_{n}}\pi^{rs}\right)\left(\Xi^{i_{m}ij}_{cpd}g^{c(a}\pi^{b)d}\left(\pi_{ij}-\frac{1}{2}g_{ij}\pi\right)\right)\right)-$$ 
\be \Xi^{i_{m}ij}_{cpd}g^{c(a}\pi^{b)d}\, \mathcal{B}^{i_{1}\cdots \hat{i}_{m}p j_{1}\cdots j_{n}}_{abrs}
\nabla_{j_{1}\cdots j_{n}}\pi^{rs}\,  \nabla_{i_{1}}\cdots\nabla_{i_{m}}\left(f(x)\left(\pi_{ij}-\frac{1}{2}g_{ij}\pi\right)\right)\label{nablapb}\ee

Using these parts of the bracket we have computed here, we will proceed
to prove our result.

\subsection{The highest derivative order bracket}
Let us recall the term we are considering: 
\be\label{equ:kinetic_original} F^{(k)}[g,\pi;x)=\mathcal{B}^{i_1\cdots i_n\,j_1\cdots  j_m}_{abcd}\left(\nabla_{i_1}\cdots\nabla_{i_n}\pi^{cd}\right)\left(\nabla_{j_1}\cdots\nabla_{j_m}\pi^{ab}\right)(x)
\ee
where $\mathcal{B}$ is a tensor depending on $g$ and $r'+2$ of its derivatives. As mentioned before, $k=r'+2+m+n$, which is the total derivative order. We are furthermore assuming that $\mathcal{B}$ imposes no contraction between $\{ab\}$ and $\{j_1\cdots j_m\}$ (resp. between $\{cd\}$ and $\{i_1\cdots i_n\}$). This is done without loss of generality, for if it were the case, we could always commute all the contracted derivatives to the right, obtaining a collection of new $\mathcal{B}$\rq{}s, with higher order derivatives of the metric and less derivatives of the momentum. One should furthermore note that, for $n=m$, we have $\mathcal{B}^{i_1\cdots i_n\,j_1\cdots  j_n}_{cdab}=\mathcal{B}^{j_1\cdots j_n\,i_1\cdots  i_n}_{abcd}$, from the then quadratic nature of $F$ in \eqref{equ:kinetic_original}. We are interested in the bracket
$$\left\{\int\textrm{d}^3x \frac{f(x)}{\sqrt{g}}F^{(k)}[g,\pi,x),\int\textrm{d}^3y \frac{h(y)}{\sqrt{g}}F_{o}[g,\pi,y)\right\}-(f\leftrightarrow h),$$
 and, we want to set one of the smearing functions $h(y)$ to a delta function, so that integration by parts is no longer allowed, and make the indices of the covariant derivatives uniform (which requires a change in dummy indices, e.g. $l\leftrightarrow i_n$). Then, we wish to isolate terms which are isolated from all the rest by virtue of containing the highest number of covariant derivatives acting on certain phase space functions. We will first display these terms and then explain in what sense they are isolated. 
 This gives us our main equation:
\be-\nabla_{i_{1}}\cdots\nabla_{i_{n}}\left(f(x)\mathcal{B}^{i_1\cdots i_n\,j_1\cdots  j_m}_{abcd}\left(\nabla_{j_1}\cdots\nabla_{j_m}\pi^{ab}\right)\right)\left (\pi^{k c}\pi_k^{~d}-\frac{1}{2}\pi\pi^{c d}\right)\label{1}\ee
\be-\nabla_{j_{1}}\cdots\nabla_{j_{m}}\left(f(x)\mathcal{B}^{i_1\cdots i_n\,j_1\cdots  j_m}_{abcd}\left(\nabla_{i_1}\cdots\nabla_{i_n}\pi^{cd}\right)\right)\left (\pi^{k a}\pi_k^{~b}-\frac{1}{2}\pi\pi^{ab}\right)\nonumber\ee
\be+(-1)^n\mathcal{B}^{i_1\cdots i_n\,j_1\cdots  j_m}_{abcd}\left(\nabla_{j_1}\cdots\nabla_{j_m}\pi^{ab}\right)\nabla_{i_{n}}\cdots\nabla_{i_{1}}\left (f(x)(\pi^{k d}\pi_k^{~c}-\frac{1}{2}\pi\pi^{cd})\right)\label{2}\ee 
\be+(-1)^m\mathcal{B}^{i_1\cdots i_n\,j_1\cdots  j_m}_{abcd}\left(\nabla_{i_1}\cdots\nabla_{i_n}\pi^{cd}\right)\nabla_{j_{m}}\cdots\nabla_{j_{1}}\left (f(x)(\pi^{k a}\pi_k^{~b}-\frac{1}{2}\pi\pi^{ab})\right)\nonumber\ee
\be+(-1)^{n}\Xi^{i_{n}ij}_{kef}g^{k(a}\pi^{b)f}\left(\pi_{ij}-\frac{1}{2}g_{ij}\pi\right) \left(\nabla_{i_{1}}\cdots\nabla_{i_{n}}\left(f(x)\mathcal{B}_{abcd}^{i_{1}\cdots \hat{i}_{n}e j_{1}\cdots j_{m}}\nabla_{j_{1}\cdots j_{m}}\pi^{cd}\right)\right)\label{3}\ee
\be+(-1)^{m}\Xi^{j_{m}ij}_{kef}g^{k(c}\pi^{d)f}\left(\pi_{ij}-\frac{1}{2}g_{ij}\pi\right) \left(\nabla_{j_{1}}\cdots\nabla_{j_{m}}\left(f(x)\mathcal{B}_{abcd}^{ i_{1}\cdots i_{n}j_{1}\cdots \hat{j}_{m}e}\nabla_{i_{1}\cdots i_{n}}\pi^{ab}\right)\right)\nonumber\ee
\be-\, \Xi^{i_{n}ij}_{kef}g^{k(c}\pi^{d)f}\, \mathcal{B}^{i_{1}\cdots \hat{i}_{n}e j_{1}\cdots j_{m}}_{abcd}
\nabla_{j_{1}\cdots j_{m}}\pi^{ab}\,  \nabla_{i_{1}}\cdots\nabla_{i_{n}}\left(f(x)\left(\pi_{ij}-\frac{1}{2}g_{ij}\pi\right)\right)\label{4}\ee
\be-\, \Xi^{j_{m}ij}_{kef}g^{k(a}\pi^{b)f}\, \mathcal{B}^{ i_{1}\cdots i_{n}j_{1}\cdots \hat{j}_{m}e}_{abcd}
\nabla_{i_{1}\cdots i_{n}}\pi^{cd}\,  \nabla_{j_{1}}\cdots\nabla_{j_{m}}\left(f(x)\left(\pi_{ij}-\frac{1}{2}g_{ij}\pi\right)\right)\nonumber\ee
\be+\sum^{r'}_{r=0}\bigg[\nabla_{l_{2}}\nabla_{l_{1}}\nabla_{a_{r}}\cdots\nabla_{a_{1}}\left({f}\frac{\partial\left(\mathcal{B}^{i_1\cdots i_n\,j_1\cdots j_m}_{abcd}\right)}{\partial(\nabla_{a_{1}}\nabla_{a_{2}}\cdots\nabla_{a_{r}}R_{kl})}\nabla_{i_1}\cdots \nabla_{i_n}\pi^{cd}\nabla_{j_1}\cdots \nabla_{j_m}\pi^{ab}\right)\left(\,^{(0)}\mathcal{A}_{kl}^{ijl_{1}l_{2}}G_{ijef}\pi^{ef}\right)\label{5}\ee
\be-(-1)^{r}\left(\frac{\partial\left(\mathcal{B}^{i_1\cdots i_n\,j_1\cdots j_m}_{abcd}\right)}{\partial(\nabla_{a_{1}}\nabla_{a_{2}}\cdots\nabla_{a_{r}}R_{kl})}\nabla_{i_1}\cdots \nabla_{i_n}\pi^{cd}\nabla_{j_1}\cdots \nabla_{j_m}\pi^{ab}\right) \nabla_{a_1}\cdots\nabla_{a_{r}}\nabla_{l_{1}}\nabla_{l_{2}}\left(f(x)\,^{(0)}\mathcal{A}_{kl}^{ijl_{1}l_{2}}G_{ijef}\pi^{ef}\right)\bigg]+\cdots\label{6} \ee

From here, we can expand in \emph{every} order of derivative of $f$, which must vanish separately. This gives us $n$ equations that should be satisfied at each space point, and for \emph{every} metric and momenta in phase space (or at least on the momentum constraint surface). It is a very tall order, and as we will see, indeed not hard to prove it can\rq{}t be done.  The problem is, as mentioned before, that we will have to divide our efforts into different subcases.

To start, we expand the two expressions which will be useful in our highest order analysis:  
   \begin{equation}\label{A}^{(0)}\mathcal{A}^{ijl_{1}l_{2}}_{kl}(\pi_{ij}-\frac{1}{2}g_{ij}\pi)=\frac{\pi}{2}(g^{l_{1}l_{2}}g_{kl}-\delta^{l_{1}}_{(k} \delta^{l_{2}}_{l)})+2\delta^{l_{2}}_{(l}\pi^{l_{1}}_{k)}-g^{l_{1}l_{2}}\pi_{kl}\end{equation}
 
     \begin{eqnarray}
    \frac{1}{2} g^{kc}\pi^{df}\Xi_{kef}^{i_{n}ij}(\pi_{ij}-\frac{1}{2}g_{ij}\pi)+\frac{1}{2} g^{kd}\pi^{cf}\Xi_{kef}^{i_{n}ij}(\pi_{ij}-\frac{1}{2}g_{ij}\pi)=\nonumber\\
   \delta^{i_{n}}_{e}\pi^{c k}\pi^{d}_{k}+\frac{1}{2}(\pi^{c i_{n}}\pi^{d}_{e}+\pi^{d i_{n}}\pi^{c}_{e})-\frac{1}{2}(g^{d i_{n}}\pi^{c k}\pi_{k e}+g^{c i_{n}}\pi^{d k}\pi_{k e})\nonumber \\ +\pi\big(-\frac{1}{2} \delta^{i_{n}}_{e}\pi^{cd}-\frac{1}{4}(g^{d i_{n}}\pi^{c}_{e}+g^{c i_{n}}\pi^{d}_{e})-\frac{1}{4}(\delta^{d}_{e} \pi^{c i_{n}} +\delta^{c}_{e} \pi^{d i_{n}})\big)\label{Xi}
     \end{eqnarray}   
     The possibly traceless part of \eqref{Xi} is:
     \begin{eqnarray}
    \delta^{i_{n}}_{e}\pi^{c k}\pi^{d}_{k}+\frac{1}{2}(\pi^{c i_{n}}\pi^{d}_{e}+\pi^{d i_{n}}\pi^{c}_{e})-\frac{1}{2}(g^{d i_{n}}\pi^{c k}\pi_{k e}+g^{c i_{n}}\pi^{d k}\pi_{k e})\label{Xi_T}
     \end{eqnarray}   
  Let us assume, without loss of generality, that $n\geq m$. 
  
  \subsubsection{Treatment of lines \eqref{5},\eqref{6}.}
  Let us consider the case when $n,m,r'$ are all non zero. In this case, the terms with the highest powers of $\nabla$ acting on momenta come from the lines \eqref{5},\eqref{6}. More specifically, \eqref{5} is the only line that contains the term 
  \be\label{ar_term}f\nabla_{l_{2}}\nabla_{l_{1}}\nabla_{a_{r\rq{}}}\cdots\nabla_{a_{1}}(\nabla_{i_1}\cdots \nabla_{i_n}\pi^{cd}\nabla_{j_1}\cdots \nabla_{j_m}\pi^{ab}),\ee
  which is a phase space function that isn't proportional to any constraint.
  
   Firstly, one should  note that even if there are contractions  between the $\left\{a_{1},\cdots, a_{r'},l_{1},l_{2}\right\}$ and $\left\{a,b,c,d\right\}$ created by $\frac{\partial \mathcal{B}_{abcd}}{\nabla_{a_{1}}\cdots\nabla_{a_{r'}}R_{kl}}$ the term won\rq{}t be annihilated on the momentum constraint surface,  because each derivative index can  contract with one momentum index, but the term is quadratic in momenta. As an example, consider the case where $\frac{\partial \mathcal{B}_{abcd}}{\nabla_{a_{1}}\cdots\nabla_{a_{r'}}R_{kl}}$ is tuned to contract any four of the $\left\{a_{i}\right\}$ indices with $\left\{a,b,c,d\right\}$, then the function takes the form:
$$\nabla_{l_1}\nabla_{l_2}\nabla_{a_{r'}}\cdots\nabla_{a_{r'-4}}\nabla_{a}\nabla_{b}\nabla_{c}\nabla_{d}(\pi^{cd}\pi^{ab})\approx \nabla_{l_1}\nabla_{l_2}\nabla_{a_{r}}\cdots\nabla_{a_{r-4}} (\nabla_{a}\nabla_{b}\pi^{cd}\nabla_{c}\nabla_{d}\pi^{ab})\neq 0.$$
Here, the $\approx$ symbol means we evaluate the left hand side on the constraint hyper surface of the momentum constraint $\nabla_{i}\pi^{ij}=0$, and we ignore terms that necessarily possess less derivatives of momenta (and more of the curvature tensor).  Similar reasoning can be applied to see why in the case when $n=0$, the function $f\nabla_{l_{2}}\nabla_{l_{1}}\nabla_{a_{r\rq{}-s}}\cdots\nabla_{a_{1}}(\pi^{cd}\nabla_{j_1}\cdots \nabla_{j_m}\pi^{ab}),$ is not wholly proportional to the diffeomorphism constraint. Moreover, no matter what sequence of commutation of covariant derivatives we perform, \eqref{ar_term} will still contain the highest order of derivatives of the momenta. 
    
        The only way the Hamiltonian constraint can remain first class then, is thus if the coefficient of \eqref{ar_term} vanishes strongly. This coefficient is
  $$\frac{\partial\left(\mathcal{B}^{i_1\cdots i_n\,j_1\cdots j_m}_{abcd}\right)}{\partial(\nabla_{a_{1}}\nabla_{a_{2}}\cdots\nabla_{a_{r\rq{}}}R_{kl})}\left(\,^{(0)}\mathcal{A}_{kl}^{ijl_{1}l_{2}}G_{ijef}\pi^{ef}\right),$$
 subsituting in $\mathcal{A}^{ijl_{1}l_{2}}_{kl}(\pi_{ij}-\frac{1}{2}g_{ij}\pi)$ from \eqref{A}, 
\be  \frac{\partial\left(\mathcal{B}^{i_1\cdots i_n\,j_1\cdots j_m}_{abcd}\right)}{\partial(\nabla_{a_{1}}\nabla_{a_{2}}\cdots\nabla_{a_{r'}}R_{kl})}\left(\frac{\pi}{2}g^{l_{1}l_{2}}g_{kl}+\delta^{l_{2}}_{(l}\pi^{l_{1}}_{k)}-g^{l_{1}l_{2}}\pi_{kl}\right) =0
\ee
The middle term has a  momentum with open indices, $l_1$,  this can\rq{}t interfere with any of the other two terms, thus  the only way for the above expression to vanish is if:
  $$\frac{\partial\left(\mathcal{B}^{i_1\cdots i_n\,j_1\cdots j_m}_{abcd}\right)}{\partial(\nabla_{a_{1}}\nabla_{a_{2}}\cdots\nabla_{a_{r\rq{}}}R_{kl})}=0.$$
  The term with the highest order of derivatives which remains after imposing the above condition is $f\nabla_{l_{2}}\nabla_{l_{1}}\nabla_{a_{r\rq{}-1}}\cdots\nabla_{a_{1}}(\nabla_{i_1}\cdots \nabla_{i_n}\pi^{cd}\nabla_{j_1}\cdots \nabla_{j_m}\pi^{ab})$, which too isn't proportional to any constraint and therefore its coefficient must vanish too. This demands:
  $$\frac{\partial\left(\mathcal{B}^{i_1\cdots i_n\,j_1\cdots j_m}_{abcd}\right)}{\partial(\nabla_{a_{1}}\nabla_{a_{2}}\cdots\nabla_{a_{r\rq{}-1}}R_{kl})}=0.$$
  Now we can iterate this reasoning with the coefficients of every successive $$f\nabla_{l_{2}}\nabla_{l_{1}}\nabla_{a_{r\rq{}-s}}\cdots\nabla_{a_{1}}(\nabla_{i_1}\cdots \nabla_{i_n}\pi^{cd}\nabla_{j_1}\cdots \nabla_{j_m}\pi^{ab})$$ for $2\leq s\leq r'$, each of which are not proportional to the constraints and thus the demand that the Hamiltonian constraint remain first class requires the strong vanishing of each of these coefficients, i.e. 
  $$\frac{\partial\left(\mathcal{B}^{i_1\cdots i_n\,j_1\cdots j_m}_{abcd}\right)}{\partial(\nabla_{a_{1}}\nabla_{a_{2}}\cdots\nabla_{a_{r\rq{}-s}}R_{kl})}=0,$$
  for $2\leq s\leq r'$. When all these conditions are imposed, there still remains the term $$f\nabla_{l_{2}}\nabla_{l_{1}}(\nabla_{i_1}\cdots \nabla_{i_n}\pi^{cd}\nabla_{j_1}\cdots \nabla_{j_m}\pi^{ab})$$ which now has the highest power of the covariant derivatives occurring and the strong vanishing of its coefficient finally requires:
  $$\frac{\partial\left(\mathcal{B}^{i_1\cdots i_n\,j_1\cdots j_m}_{abcd}\right)}{\partial R_{kl}}=0.$$
 Thus the first class nature of the Hamiltonian constraint requires $\mathcal{B}^{\left\{i\right\}\left\{j\right\}}_{abcd}$ to depend on the metric alone and none of its derivatives. In other words it must be ultra local in the metric. 
 \ \
 
 Notice that the above analysis pertains only to terms appearing in lines \eqref{5}, \eqref{6} of our main equation. When all the above conditions are imposed, all terms in \eqref{5} and \eqref{6} vanish. We can afford to restrict our attention to just these lines because the highest power of the covariant derivatives which appears in the other terms in the main equation contain at most $m+n$ derivatives whereas the terms in \eqref{5},\eqref{6} contain at the very least $m+n+2$ powers of the `loose' covariant derivatives acting on momenta and smearing functions. Thus there is no room for interference from the other lines.
 \ \
 
 We note that the above proof applies equally well when either or both of $m,n$ are zero. In this case, the phase space functions we will choose take respectively the form 
 $$f\nabla_{l_{2}}\nabla_{l_{1}}\nabla_{a_{r\rq{}-s}}\cdots\nabla_{a_{1}}(\pi^{cd}\nabla_{j_1}\cdots \nabla_{j_m}\pi^{ab}),$$
 and
 $$f\nabla_{l_{2}}\nabla_{l_{1}}\nabla_{a_{r\rq{}-s}}\cdots\nabla_{a_{1}}(\pi^{cd}\pi^{ab}),$$
 for $0\leq s\leq r'$. The demand that the coefficients of these terms vanish will lead to the same conditions as the ones we have above, i.e. to the vanishing of each $\frac{\partial\left(\mathcal{B}^{j_1\cdots j_m}_{abcd}\right)}{\partial(\nabla_{a_{1}}\nabla_{a_{2}}\cdots\nabla_{a_{r\rq{}-s}}R_{kl})}$ and of each $\frac{\partial\left(\mathcal{B}_{abcd}\right)}{\partial(\nabla_{a_{1}}\nabla_{a_{2}}\cdots\nabla_{a_{r\rq{}-s}}R_{kl})}$ for $0\leq s\leq r'$.  Henceforth we shall proceed considering only the ultra local $\mathcal{B}^{\left\{i\right\}\left\{j\right\}}_{abcd}$, in other words, in all cases that follow, $r'=0$. 
$$\square$$

\subsection{Treatment of the remaining terms in the bracket.}
Now we can explain the reason behind choosing the terms that we have for the main equation coming form computing the Poisson bracket.

From the above case we see that the highest order term in the sum in lines \eqref{5},\eqref{6} which is clearly isolated by virtue of the fact that it contains $m+n+r'$ powers of the covariant derivative operator, and from the computation in \eqref{rterm} we see that there is no other source for such a term and thus no room  for interference. The reason why we retained a sum of all terms containing $\, ^{(0)}\mathcal{A}_{kl}^{ijl_{1}l_{2}}G_{ijef}\pi^{ef}$ is because when we see that the vanishing of $\frac{\partial \mathcal{B}^{\left\{i\right\}\left\{j\right\}}_{abcd}}{\partial \nabla_{\left\{r\right\}}R_{kl}}$ will kill all terms it multiplies (i.e. those involving the $\, ^{(n)}\mathcal{A}_{a_{1}\cdots a_{n}kl}^{ijl'}$ that depends on the curvature tensor and such but lower powers of $\nabla$) so the term containing the highest power of the covariant derivatives among those that remain will be the last but one term in the sum, i.e. the one containing $m+n+r'-1$ powers of $\nabla$. This way we keep descending until we reach the lowest order term containing $m+n+2$ powers of derivatives and as we saw above these terms too must vanish in order for the constraint to be first class. 

Now we will have to seek the terms with the highest powers the covariant derivatives among the rest of the terms generated by the Poisson bracket and thus these terms will have to contain $m+n$ derivatives. Among these are the terms in the lines \eqref{1}-\eqref{4}. From the computations done leading to \eqref{nablapb}, we also see that these are isolated in the sense that all other terms are somehow or other sub-leading to these terms, despite having perhaps the same power of the covariant derivatives occurring. This is because even among terms containing $m+n$ occurrences of the covariant derivatives in total, there is still a hierarchy among terms with different number of derivatives hitting different phase space functions such a the momentum or the smearing function. The terms we have chosen in lines \eqref{1}-\eqref{4} are isolated because they are of highest order in this hierarchy among terms containing $m+n$ derivatives. 

  \subsubsection{Case: $n\geq m \neq0$}

    \subsubsection*{Odd $n\geq m$ }
   
  This part of the proof works even for $m=0$. 
    
  For $n$ odd, $(-1)^n=-1$. Then for the equation representing the coefficient of $\nabla_{i_1}\cdots \nabla_{i_n} f(x)$, the first line of \eqref{1} is identical to the first line of \eqref{2}, after commutation (which only produces terms that are sub-leading in derivatives of $f(x)$) and the first line of \eqref{3} is identical to the first line of \eqref{4}. The same argument can be repeated, mutatis mutandis, for the second lines. 

Then we get  the equation representing the coefficient of
 $(\nabla_{j_1}\cdots \nabla_{j_m}\pi^{ab})(\nabla_{i_1}\cdots \nabla_{i_n}f(x))$ (if $n=m$, we  do this after trading indices twice for the second lines of \eqref{1} and \eqref{3}, which doesn\rq{}t alter the overall product),  
\begin{eqnarray*}
{\mathcal{B}}\,^{i_1\cdots i_n\,j_1\cdots  j_m}_{abcd}\left (\pi^{k c}\pi_k^{~d}-\frac{1}{2}\pi\pi^{c d}\right)
+\Xi_{kef}^{i_{n}ij}g^{k(c}\pi^{d)f}(\pi_{ij}-\frac{1}{2}\pi g_{ij})\mathcal{B}^{i_{1}\cdots \hat{i}_{n}e\,j_1\cdots  j_m}_{abcd}=0
\end{eqnarray*}
From \eqref{Xi_T} we have that the potentially traceless parts of  $\Xi_{kef}^{i_{n}ij}g^{k(c}\pi^{d)f}(\pi_{ij}-\frac{1}{2}\pi g_{ij})\mathcal{B}^{i_{1}\cdots \hat{i}_{n}e\,j_1\cdots  j_m}_{abcd}$ is of the form: 
\begin{eqnarray}\mathcal{B}^{i_{1}\cdots \hat{i}_{n}e\,j_1\cdots  j_m}_{abcd}\left(  \delta^{i_{l}}_{e}\pi^{c k}\pi^{d}_{k}+\pi^{ i_{l}(c}\pi^{d)}_{e}-g^{ i_{l}(d}\pi^{c) k}\pi_{k e}\right)= {\mathcal{B}}\,^{i_1\cdots i_n\,j_1\cdots  j_m}_{abcd} \pi^{c k}\pi^{d}_{k}
\nonumber\\
+\mathcal{B}^{i_{1}\cdots \hat{i}_{n}e\,j_1\cdots  j_m}_{abcd}\left(\pi^{ i_{l}(c}\pi^{d)}_{e}-g^{ i_{l}(d}\pi^{c) k}\pi_{k e}\right)\label{whatever}
\end{eqnarray}
Inputting \eqref{Xi_T}, for the possibly traceless part: 
\begin{eqnarray}\label{m=n=odd}
2\mathcal{B}\,^{i_1\cdots i_n\,j_1\cdots  j_m}_{abcd}\pi^{k c}\pi_k^{~d}
+\mathcal{B}^{i_{1}\cdots \hat{i}_{n}e\,j_1\cdots  j_m}_{abcd}\left(\pi^{ i_{n}(c}\pi^{d)}_{e}-g^{ i_{n}(d}\pi^{c)k}\pi_{k e}\right)=0
\end{eqnarray}
Note that these terms cannot interfere with those of \eqref{5} and \eqref{6}, since in those cases we do not have a term which leaves two momenta undifferentiated. 

Now, $i_n$ is an open index in \eqref{m=n=odd}. For the first element of the equation to have such an index, we would need to have 
\be\label{forbidden}  {\mathcal{B}}\,^{i_1\cdots i_n\,j_1\cdots  j_m}_{abcd}=  \alpha\mathcal{A}^{i_{1}\cdots \hat{i}_{n}\,j_1\cdots  j_m}_{ab(c}\delta^{i_n}_{d)}+\beta  \mathcal{C}^{i_1\cdots i_{n}\,j_1\cdots  j_m}_{ab}g_{cd}
\ee

But the $\alpha$ terms all break our clause that there should be no such types of contractions, e.g. $\delta^{i_n}_d$ (or $\delta^{j_m}_b$). This sets $\alpha=0$, and thus the first term of \eqref{m=n=odd}  cannot have coefficients to cancel the second term. The second and third term could cancel by themselves however. That case corresponds to   
\be\label{propto_metric}  {\mathcal{B}}\,^{i_1\cdots i_n\,j_1\cdots  j_m}_{abcd}=  \mathcal{C}^{i_1\cdots i_{n}\,j_1\cdots  j_m}_{ab}g_{cd}\ee
which is also the required form of $\mathcal{B}$ if the last component is to have a free $i_n$ index on the momenta.    Furthermore, for such terms,  note that 
$$ g_{cd}\left(\pi^{ j_{l}(c}\pi^{d)}_{e}-g^{ j_{l}(d}\pi^{c) k}\pi_{k e}\right)=0
$$
leaving the first component of \eqref{m=n=odd}, 
$$2  \mathcal{C}^{i_1\cdots i_{n}\,j_1\cdots  j_m}_{ab}\pi^{kc}\pi_{kc}=0$$
 finishing this part of the proof.  

 \subsubsection*{Even $n\geq m>0$} 
 Now we move forward with the even case for the highest number of derivatives, $n$. For this, we will split into subcases, i) ${\mathcal{B}}\,^{i_1\cdots i_n\,j_1\cdots  j_m}_{abcd}\propto g_{ab}$ or  ${\mathcal{B}}\,^{i_1\cdots i_n\,j_1\cdots  j_m}_{abcd}\propto g_{cd}$, or ii) ${\mathcal{B}}\,^{i_1\cdots i_n\,j_1\cdots  j_m}_{abcd}\not\propto g_{ab}$ and ${\mathcal{B}}\,^{i_1\cdots i_n\,j_1\cdots  j_m}_{abcd}\not\propto g_{cd}$.  Let\rq{}s start with the second case. 

Here we will take the coefficient of $f(x)\nabla_{j_{1}\cdots j_{m}}\pi^{ab}\,\nabla_{i_1}\cdots \nabla_{i_n}\pi$ without derivatives on a third momentum. These come from the first line of \eqref{2} and the first line of \eqref{4} for $n>m$, and from both lines of \eqref{2} and \eqref{4} for $n=m$. In the latter case, we just get twice the first lines, from ${\mathcal{B}}\,^{i_1\cdots i_n\,j_1\cdots  j_n}_{abcd}\leftrightarrow {\mathcal{B}}\,^{j_1\cdots j_n\,i_1\cdots  i_n}_{cdab}$.  Note furthermore from \eqref{Xi_uncontracted} that
$$\Xi_{kab}^{lij}\pi_{ij}=-\delta^l_k\pi_{ab}+\delta^l_b\pi_{ak}+\delta_a^l\pi_{bk}
$$
thus there are no trace terms coming from $\pi_{ij}$ inside the derivatives, and the derivative of the trace of the momentum from \eqref{4} comes only from $\Xi_{kab}^{lij}\, g_{ij}\pi$, and the only coefficients of $f(x)\nabla_{j_{1}\cdots j_{m}}\pi^{ab}\,\nabla_{i_1}\cdots \nabla_{i_n}\pi$ 
\be\label{not_propto}
- \mathcal{B}^{i_1\cdots i_n\,j_1\cdots  j_m}_{abcd}\pi^{cd}
 +\, \Xi^{i_{n}ij}_{kef}g^{k(c}\pi^{d)f}\, \mathcal{B}^{i_{1}\cdots \hat{i}_{n}e j_{1}\cdots j_{m}}_{abcd}
g_{ij}=0
\ee
Now, 
\be\label{Xi_contracted}\Xi^{i_nij}_{kef}g_{ij}g^{k(c}\pi^{d)f}= \delta^{i_n}_e\pi^{cd}-2g^{i_n(c}\pi^{d)}_e+2\delta^{(c}_e\pi^{d)i_n}\ee
thus we are left with 
$$\mathcal{B}^{i_{1}\cdots \hat{i}_{n}e j_{1}\cdots j_{m}}_{abcd}(-2g^{i_n(c}\pi^{d)}_e+2\delta^{(c}_e\pi^{d)i_n})=0$$
Again, $i_n$ is an open index. For the first term to have a momentum with such an index, we would need  ${\mathcal{B}}\,^{i_1\cdots i_n\,j_1\cdots  j_m}_{abcd}\propto g_{cd}$, but that requires us to consider other types of terms, and lies beyond case ii. 

Moving on to case i, suppose that  ${\mathcal{B}}\,^{i_1\cdots i_n\,j_1\cdots  j_m}_{abcd}\propto g_{cd}$ but  ${\mathcal{B}}\,^{i_1\cdots i_n\,j_1\cdots  j_m}_{abcd}\not\propto g_{ab}$.  In other words, as in \eqref{propto_metric}, 
$$ {\mathcal{B}}\,^{i_1\cdots i_n\,j_1\cdots  j_m}_{abcd}= \mathcal{C}^{i_1\cdots i_{n}\,j_1\cdots  j_m}_{ab}g_{cd}$$ with $\mathcal{C}^{i_1\cdots i_{n}\,j_1\cdots  j_m}_{ab}\not\propto g_{ab}$.\footnote{One should note that this could still include  $\mathcal{C}^{i_1\cdots i_{n}\,j_1\cdots  j_m}_{ab}=\alpha\mathcal{D}^{i_1\cdots i_{n}\,j_1\cdots  j_m}g_{ab}+\beta \widetilde{\mathcal{C}}^{i_1\cdots i_{n}\,j_1\cdots  j_m}_{ab}$ for $\widetilde{\mathcal{C}}^{i_1\cdots i_{n}\,j_1\cdots  j_m}_{ab}\not\propto g_{ab}$ as long as $\beta\neq 0$. The point is still that the calculations involving $\beta$ would leave terms which cannot interfere with the $\alpha$ terms.  This is always what we mean with statements of the sort ${\mathcal{B}}\,^{i_1\cdots i_n\,j_1\cdots  j_m}_{abcd}\not\propto g_{ab}$.} 
 Here, only the second line of \eqref{2}, the second line of \eqref{4}, and possibly \eqref{6} can contribute with a term such as $\nabla_{\{i\}}\pi \nabla_{\{j\}}\pi$. However, in the case of \eqref{6}, the term would also include derivatives of a third momentum, unlike \eqref{2} and \eqref{4}. Again, from \eqref{Xi_uncontracted}, we need only worry about the trace term in \eqref{4}. We thus obtain, for the coefficient of $f(x)\nabla_{j_{1}\cdots j_{m}}\pi\,\nabla_{i_1}\cdots \nabla_{i_n}\pi$,
 \begin{align}  (-1)^m\pi^{ab}\mathcal{C}^{i_1\cdots i_n\,j_1\cdots  j_m}_{ab}-\, g_{ij}\Xi^{j_{m}ij}_{kef}g^{k(a}\pi^{b)f}\, \mathcal{C}^{ i_{1}\cdots i_{n}j_{1}\cdots {j}_{m-1}e}_{ab}\nonumber\\
 =( (-1)^m-1)\pi^{ab}\mathcal{C}^{i_1\cdots i_n\,j_1\cdots  j_m}_{ab}+(-2g^{j_m(a}\pi^{b)}_e+2\delta^{(a}_e\pi^{b)j_m})\mathcal{C}^{ i_{1}\cdots i_{n}j_{1}\cdots {j}_{m-1}e}_{ab}=0
\label{almost} \end{align}
where we used \eqref{Xi_contracted}. If $m$ is even, the first term of \eqref{almost} vanishes. Trying to obtain the $j_m$ open index for the momentum in the middle term,  we would reach the same conclusion as in case ii, namely, that $\mathcal{C}^{i_1\cdots i_{n}\,j_1\cdots  j_m}_{ab}\propto g_{ab}$, which also lies beyond this subcase, and will be considered separately. If $m$ is odd, in principle we could try to obtain that momentum with the open index from the first term of \eqref{almost}, uniquely through 
$$ \mathcal{C}^{i_1\cdots i_{n}\,j_1\cdots  j_m}_{ab}=\mathcal{D}_{(a}^{i_1\cdots i_{n}\,j_1\cdots  j_{m-1}}\delta ^{j_m}_{b)}$$
Inputting this back on \eqref{almost}, we get: 
$$-\mathcal{D}_{a}^{i_1\cdots i_{n}\,j_1\cdots  j_{m-1}}g^{ j_m\, a}\pi+3 \mathcal{D}_{a}^{i_1\cdots i_{n}\,j_1\cdots  j_{m-1}}\pi^{j_m a}=0$$
and here there is nothing more we can tune to make the trace term proportional to the full momentum with an open index, except setting $\mathcal{D}_{a}^{i_1\cdots i_{n}\,j_1\cdots  j_{m-1}}=0$. We would obtain the same result as this last part if we chose ${\mathcal{B}}\,^{i_1\cdots i_n\,j_1\cdots  j_m}_{abcd}\not\propto g_{cd}$ but  ${\mathcal{B}}\,^{i_1\cdots i_n\,j_1\cdots  j_m}_{abcd}\propto g_{ab}$, since we know that $n$ is even (and otherwise the two cases are identical upon an exchange of $m\leftrightarrow n, ab\leftrightarrow cd$). 

Finally, we move on to the subcase where 
\be\label{reduced_B} {\mathcal{B}}\,^{i_1\cdots i_n\,j_1\cdots  j_m}_{abcd}= \mathcal{C}^{i_1\cdots i_{n}\,j_1\cdots  j_m}g_{ab}g_{cd}\ee
For this case, we will look at coefficients of $f(x)\pi^{cd}\nabla_{j_{1}\cdots j_{m}}\pi\,\nabla_{i_1}\cdots \nabla_{i_n}\pi_{cd}$. These can come solely from \eqref{2} and \eqref{4}. For \eqref{reduced_B} we have:
 \be\label{Xi_more_contracted}
 \Xi^{i_nij}_{kef}g^{k(c}\pi^{d)f}{\mathcal{B}}\,^{i_1\cdots i_{n-1}e\,j_1\cdots  j_m}_{abcd}=\mathcal{C}^{i_1\cdots i_{n}\,j_1\cdots  j_m}g_{ab}\, \pi^{ij} 
 \ee
Now note that \eqref{2} will get twice the contribution of $\pi^{cd}\,\nabla_{i_1}\cdots \nabla_{i_n}\pi_{cd}$ since the derivatives are acting on the square $\pi^{cd}\pi_{cd}$. Putting this back in and summing with \eqref{4} we get, for the coefficient of $f(x)\pi^{cd}\nabla_{j_{1}\cdots j_{m}}\pi\,\nabla_{i_1}\cdots \nabla_{i_n}\pi_{cd}$ simply:
\be \mathcal{C}^{i_1\cdots i_{n}\,j_1\cdots  j_m}=0\label{everything}\ee
without further calculation.
$$\square$$ 
 
 This establishes all the cases for which $n\geq m>0$. Now we move on to the case $m=0$.
 
 \subsubsection{Case: $n\geq m=0$}
 For this case, one should first of all notice that the second line of \eqref{1} cancels with the second line of \eqref{2}, and the second line of \eqref{3} cancels with the second line of \eqref{4}. 
 
  We will take the coefficient of $f(x)\pi^{ab}\pi^{cd}\nabla_{i_1}\cdots\nabla_{i_n}\pi$. Again, we split it into two cases i) ${\mathcal{B}}\,^{i_1\cdots i_n}_{abcd}\propto g_{cd}$ and/or ${\mathcal{B}}\,^{i_1\cdots i_n}_{abcd}\propto g_{ab}$ and ii) ${\mathcal{B}}\,^{i_1\cdots i_n}_{abcd}\not\propto g_{cd}$ and ${\mathcal{B}}\,^{i_1\cdots i_n}_{abcd}\not\propto g_{ab}$. Let\rq{}s start with the first case. 

  In this case, the only terms that contribute come from the first  line of \eqref{2} and the first line of \eqref{4}. In this case, despite the fact that line \eqref{5} would also contain two undifferentiated momenta, it wouldn't contribute towards to coefficient of $f(x)\pi^{ab}\pi^{cd}\nabla_{i_1}\cdots\nabla_{i_n}\pi$ because it necessarily would contain $n+r+2$ derivatives leading to a term of the form $f(x)\pi^{ab}\pi^{cd}\nabla_{a_{1}}\cdots\nabla_{a_{r}}\nabla_{l_{1}}\nabla_{l_{2}}\nabla_{i_1}\cdots\nabla_{i_n}\pi$, and thus contains at least two more derivatives than the term we are interested in. As in \eqref{not_propto},  the derivative of the trace of the momentum from \eqref{4} comes only from $\Xi_{kab}^{lij}\, g_{ij}\pi$, and thus the only coefficients of $f(x)\pi^{ab}\,\nabla_{i_1}\cdots \nabla_{i_n}\pi$ which are linear in the momentum are 
\be
- \mathcal{B}^{i_1\cdots i_n}_{abcd}\pi^{cd}
 +\, \Xi^{i_{n}ij}_{kef}g^{k(c}\pi^{d)f}\, \mathcal{B}^{i_{1}\cdots \hat{i}_{n}e}_{abcd}
g_{ij}=0
\ee
which is exactly the same as \eqref{not_propto}. This concludes this case. 

Moving on to the case  ${\mathcal{B}}\,^{i_1\cdots i_n}_{abcd}\propto g_{cd}$ and ${\mathcal{B}}\,^{i_1\cdots i_n}_{abcd}\not\propto g_{ab}$, we will focus on coefficients of terms of the form: $f(x)\pi^a_{~c}\pi^{bc}\nabla_{i_1}\cdots \nabla_{i_n}\pi$. 
   Lines \eqref{1} and \eqref{2} don\rq{}t contain such terms, only \eqref{3} and \eqref{4}. For the coefficients of  $f(x)\nabla_{i_1}\cdots \nabla_{i_n}\pi$ which can contain the trace-free part of the momenta we obtain at first, 
   \be \left(\Xi^{i_{n}ij}_{kef}g^{k(a}\pi^{b)f}\left(\pi_{ij}-\frac{1}{2}g_{ij}\pi\right)+\frac12\, \Xi^{i_{n}ij}_{kef}g^{k(c}\pi^{d)f}\, 
\pi^{ab}\, g_{ij} g_{cd}\right)\mathcal{C}_{ab}^{i_{1}\cdots \hat{i}_{n}e}\ee
     but from \eqref{Xi_contracted}, 
     \be\Xi^{i_nij}_{kef}g_{ij}g^{k(c}\pi^{d)f}g_{cd}= \delta^{i_n}_e\pi\ee
     and from    \eqref{Xi_T}, the possibly  traceless part of $ g^{k(a}\pi^{b)f}\Xi_{kef}^{i_{n}ij}(\pi_{ij}-\frac{1}{2}g_{ij}\pi)$ gives:        
     \be \mathcal{C}_{ab}^{i_{1}\cdots {i}_{n}}\pi^{a k}\pi^{b}_{k}+\left(\pi^{ i_{n}(a}\pi^{b)}_{e}-g^{ i_{n}(a}\pi^{b) k}\pi_{k e}\right)\mathcal{C}_{ab}^{i_{1}\cdots \hat{i}_{n}e}=0\ee
this is of the same form as \eqref{whatever}, and the same analysis as applied to \eqref{m=n=odd} follows. 

Now we analyze the case ${\mathcal{B}}\,^{i_1\cdots i_n}_{abcd}\not\propto g_{cd}$ and ${\mathcal{B}}\,^{i_1\cdots i_n}_{abcd}\propto g_{ab}$. For this, we will look again at the terms of the form $f(x)\pi^{c k}\pi^{d}_{k}\nabla_{i_1}\cdots \nabla_{i_n}\pi$, i.e. with two contracted undifferentiated momenta. This comes solely from line \eqref{1}. Line \eqref{2} won\rq{}t contribute, because the term multiplying  $f(x)\nabla_{i_1}\cdots \nabla_{i_n}\pi$ will contain a trace of the momentum. The same happens with line \eqref{4}, and line \eqref{3} won\rq{}t contain a term like $f(x)\nabla_{i_1}\cdots \nabla_{i_n}\pi$. This results immediately on $ \mathcal{C}_{cd}^{i_{1}\cdots {i}_{n}}=0$. 

Finally, we discard the possibility that ${\mathcal{B}}\,^{i_1\cdots i_n}_{abcd}\propto g_{ab}g_{cd}$. Considering the term $f(x)\pi\pi^{cd}\,\nabla_{i_1}\cdots \nabla_{i_n}\pi_{cd}$, this case is identical to the one for $m\neq 0$, in equations \eqref{reduced_B} to \eqref{everything}.  This concludes the proof. $\square$.

\section{Can Closure Protect Reversibility?}
The modifications considered in the previous section were still predicated on two powers of the momentum in the Hamiltonian constraint. This condition is consequential for time reversal invariance of the theory. If we were to consider in addition to generalisations of the kinetic term which are quadratic in the momentum, an additional term linear in the momentum, then the resulting theory could potentially break time reversal invariance. In this section we will consider a generalised Hamiltonian constraint of the form
$$H(N)=\int \textrm{d}^{3}xN(x)(\Phi[g,\pi;x)+\sqrt{g}\hat{V}[g,\pi;x)+P[g,\pi;x)),$$
 where
 \be 
 P[g,\pi;x)=-\beta[g;x)_{ab}\pi^{ab}.
 \ee
The tensor $\beta^{ab}[g;x)$ is a local function of the metric and its derivatives up to arbitrary order $s$. In what is to follow, we will not compute the relevant brackets with the use of smearing functions and shall deal directly with the resulting distributions themselves. 
Let us compute the $\mathcal{O}(\pi^{2})$ part of the bracket,
$$ \{   G_{abcd}(x) \pi^{ab}\pi^{cd}(x)+\beta_{ab}(x)\pi^{ab}(x)\,,   G_{ijkl}(y) \pi^{ij}\pi^{kl}(y)+\beta_{ij}(y)\pi^{ij}(y)\},
$$
whose non-zero part may only come from the bracket:
$$\{   G_{abcd}(x) \pi^{ab}\pi^{cd}(x),\beta_{ij}(y)\pi^{ij}(y)\}-(x\leftrightarrow y).
$$
This leads to the expression 
$$ \pi^{ab}(x)G_{abcd}(x)\frac{\delta \beta_{ef}(y)}{\delta g_{cd}(x)}\pi^{ef}(y)-(x\leftrightarrow y),$$
and we can rewrite this as
$$ \pi^{ab}(x)\pi^{ef}(y)\left(G_{abcd}(x)\frac{\delta\beta_{ef}(y)}{\delta g_{cd}(x)}-G_{efcd}(y)\frac{\delta \beta_{ab}(x)}{\delta g_{cd}(y)}\right).$$
This has to vanish weakly for the constraint algebra to be first class. Because this term is quadratic in the momenta, one possibility is that the kinetic term of the Hamiltonian constraint might be formed from this bracket. 

The distribution $\frac{\delta\beta_{ab}(x)}{\delta g_{ecd}(y)}$ containing derivatives of the delta function will not do the trick because these derivatives inadvertently will end up acting on only one of the momenta, and never on both. However, if  the bracket is to result in some structure functions contracting with derivatives of the Hamiltonian constraint, we would need derivatives acting on both momenta. Discarding possible derivatives of the $\delta(x,y)$, we see that for proportionality with the Hamiltonian constraint we must have $\frac{\delta\beta_{ef}(y)}{\delta g_{cd}(x)}=F[g;x)\delta^{cd}_{ef}\delta(x,y),$ and $\frac{\delta \beta_{ab}(x)}{\delta g_{cd}(y)}=F[g;x)\delta^{cd}_{ab}\delta(x,y)$.  This case is trivial however, as this choice will make the bracket vanish due to the even property of $\delta(x,y)$ under the exchange $(x\leftrightarrow y)$.  Thus we cannot form the Hamiltonian constraint from the brackets involving the term $\beta_{ab}\pi^{ab}$ at all. 

Now to consider the diffeomorphism constraint. All we need to show here is that \emph{all} the terms in $\frac{\delta\beta_{ab}(x)}{\delta g_{cd}(y)}$ do not, or cannot, contain a covariant divergence of the momentum tensor. First, let us note that at every order $r$ of derivatives of the delta function in the expansion of $\frac{\delta\beta_{ab}(x)}{\delta g_{cd}(y)}$, there are terms of the form
$$ G_{cdij}\pi^{ij}(x)\pi^{ab}(y)\frac{\delta\beta_{ab}(x)}{\delta g_{cd}(y)}\supset G_{cdij}\pi^{ij}(x)\pi^{ab}(y)\frac{\partial \beta_{ab}}{\partial(\nabla_{a_{1}}\cdots \nabla_{a_{r-2}}R_{kl})}\, ^{(0)}\mathcal{A}^{cdl_{1}l_{2}}_{kl}\nabla^{x}_{a_{1}}\cdots\nabla^{x}_{a_{r-2}}\nabla^{x}_{l_{1}}\nabla^{x}_{l_{2}}\delta(x,y).$$
Then, we note that  if in principle we choose to smear the above expression with some function over space, the derivatives will all hit $\, ^{(0)}\mathcal{A}^{cdl_{1}l_{2}}_{kl}G_{cdij}\pi^{ij}(x)$. Due to the presence of the trace term, i.e. $\, ^{(0)}\mathcal{A}^{cdl_{1}l_{2}}_{kl}G_{cdij}\pi^{ij}(x)\supset g^{l_{1}l_{2}}\pi(x)$, whose derivative would be non zero, we find that there are always terms in $\frac{\delta\beta_{ab}(x)}{\delta g_{cd}(y)}$ which cannot be proportional to the diffeomorphism constraint without imposing further constraints, no matter what the form of $\frac{\partial \beta_{ab}}{\partial(\nabla_{a_{1}}\cdots \nabla_{a_{r-2}}R_{kl})}$. Also note that the only situation where all the terms at every order of derivatives of the delta function in $\frac{\partial \beta_{ab}}{\partial(\nabla_{a_{1}}\cdots \nabla_{a_{r-2}}R_{kl})}$ somehow vanish is when $\beta_{ab}\propto g_{ab}$ in which case the linear term trivially commutes with the kinetic term. 
Thus we conclude that the strong vanishing of the bracket of interest is the only way to ensure the first class nature of the Hamiltonian constraint. 

The strong vanishing of this bracket then implies that
\be \pi^{ab}(x)\pi^{ef}(y)\left(G_{abcd}(x)\frac{\delta\beta_{ef}(y)}{\delta g_{cd}(x)}-G_{efcd}(y)\frac{\delta \beta_{ab}(x)}{\delta g_{cd}(y)}\right)=0.\ee
We would like to first rewrite the terms inside the round brackets as a difference between total functional derivatives. This can be done by rewriting the above expression by inserting the identity as
$$\delta^{ab}_{rs}\pi^{rs}(x)\delta^{ef}_{pq}\pi^{pq}(y)\left(G_{abcd}(x)\frac{\delta\beta_{ef}(y)}{\delta g_{cd}(x)}-G_{efcd}(y)\frac{\delta \beta_{ab}(x)}{\delta g_{cd}(y)}\right),$$
then, when we write $\delta^{ab}_{rs}=G^{abij}(x)G_{ijrs}(x)$ and $\delta^{ef}_{pq}=G^{efkl}(y)G_{klpq}(y)$, we can write
\be\label{identity_insert}G_{ijrs}(x)\pi^{rs}(x)G_{klpq}(y)\pi^{pq}(y)\left(G^{efkl}(y)\frac{\delta\beta_{ef}(y)}{\delta g_{ij}(x)}-G^{abij}(x)\frac{\delta \beta_{ab}(x)}{\delta g_{kl}(y)}\right).\ee
We can now write the terms inside the round brackets as 
$$G^{efkl}(y)\frac{\delta\beta_{ef}(y)}{\delta g_{ij}(x)}-G^{abij}(x)\frac{\delta \beta_{ab}(x)}{\delta g_{kl}(y)}=\left(\frac{\delta (G^{efkl}(y)\beta_{ef}(y))}{\delta g_{ij}(x)}-\frac{\delta (G^{abij}(x)\beta_{ab}(x))}{\delta g_{kl}(y)}\right)+
$$
$$+\left(\frac{\delta G^{efkl}(y)}{\delta g_{ij}(x)}\beta_{ef}(y)-\frac{\delta G^{abij}(x)}{\delta g_{kl}(y)}\beta_{ab}(x)\right).$$
For notational convenience we will define $G_{abcd}\pi^{cd}\equiv \tilde{K}_{ab}$.
The second line will vanish by symmetrization properties of the indices and the undifferentiated $\delta$ function. It is also straightforward to compute, we find that it yields:
$$\left(\frac{\delta G^{efkl}(y)}{\delta g_{ij}(x)}\beta_{ef}(y)-\frac{\delta G^{abij}(x)}{\delta g_{kl}(y)}\beta_{ab}(x)\right)=\left(\beta^{ik}(x)g^{jl}(x)-\beta^{il}(x)g^{jk}(x)\right)\delta(x,y),$$
where we have used the properties of the delta function to write this in a suitable form. When we plug this back into the bracket we were computing, we find 
$$\left(\beta^{il}\tilde{K}_{ij}\tilde{K}^{j}_{l}-\beta^{jk}\tilde{K}_{kl}\tilde{K}^{l}_{j}\right)(x)\delta(x,y)+$$
$$\tilde{K}_{ij}(x)\tilde{K}_{kl}(y)\left(\frac{\delta (G^{efkl}(y)\beta_{ef}(y))}{\delta g_{ij}(x)}-\frac{\delta (G^{abij}(x)\beta_{ab}(x))}{\delta g_{kl}(y)}\right).$$
The first line vanishes, and in order for the second line to vanish, we are left with the condition
\begin{equation} \label{curlfree} \frac{\delta (G^{efkl}(y)\beta_{ef}(y))}{\delta g_{ij}(x)}-\frac{\delta (G^{abij}(x)\beta_{ab}(x))}{\delta g_{kl}(y)}=0 \end{equation}

Now, we can consider the bracket 
$$
\left\{\int \textrm{d}^{3}x f(x)\left(\beta_{ab}\pi^{ab}(x)+\sqrt{g}\hat{V}[g;x)\right),\int\textrm{d}^{3}y h(y)\left(\beta_{ab}\pi^{ab}(y)+\sqrt{g}\hat{V}[g;y)\right)\right\}.
$$
Computing this bracket yields the expression
$$
\int \textrm{d}^{3}z \int \textrm{d}^{3}x h(z)\left(\beta_{ab}(z)\frac{\delta \hat{V}[g;x)}{\delta g_{ab}(z)}\right)f(x)-(z\leftrightarrow x),
$$
The fact that the bracket between the kinetic term and this linear term vanishes dashes any hopes to form the Hamiltonian constraint out of the bracket between the Hamiltonian constraint and itself. Given that the above bracket is independent of the momenta, it too has to strictly vanish, because the bracket between two Hamiltonian constraints is now forced to be proportional to the diffeomoprhism constraint in order that it remain first class. 
The distribution $\left(\frac{\delta \hat{V}[g;x)}{\delta g_{ab}(z)}\right)$ splits into a sum of terms which we can write as 
$$
\left(\beta_{ab}(z)\frac{\delta \hat{V}[g;x)}{\delta g_{ab}(z)}\right)=\beta_{ab}(z)\sum_{n}f^{abi_{1}\cdots i_{n}}[g;x)\nabla_{i_{1}}\cdots\nabla_{i_{n}}\delta(x,z).
$$
The strong vanishing of the above bracket implies the vanishing of terms at every derivative order. The lowest order term proportional to $\delta(x,z)$ vanishes due to the exchange $(x\leftrightarrow z)$. This is not as simple as demanding that each of the $f^{abi_{1}\cdots i_{n}}[g;x)$ vanish because they are not all of order zero, but they too contain terms of different derivative orders. For instance, the next to leading order contains the lowest order part of $f^{abi_{1}i_{2}}[g;x)$ which we will call $f^{abi_{1}i_{2}}_{(0)}[g;x)$ and this term will be independent of derivatives of the metric. From straightforward computation, we find that 
$$
f^{abi_{1}i_{2}}[g;x)\supset \frac{\partial \hat{V}}{\partial R_{kl}}\, ^{(0)}\mathcal{A}^{ab i_{1}i_{2}}_{kl},
$$
now, the lowest order part $f^{abi_{1}i_{2}}_{(0)}$ corresponds to the term $g_{kl}\subset \frac{\partial V}{\partial R_{kl}}.$ Thus we see that this term comes from the bracket between the linear term and the second order part of the potential which contains the Ricci tensor. So we find that 
$$
f^{abi_{1}i_{2}}_{(0)}=G^{abi_{1}i_{2}}.
$$
We then see that the Poisson bracket of interest at this order yields: 
$$
\int \textrm{d}^{3}z \int\textrm{d}^{3}x f(x)h(z)\left(G^{ab i_{1}i_{2}}\beta_{ab}(z)\nabla^{z}_{i_{1}}\nabla^{z}_{i_{2}}\delta(x,z)\right)-(x\leftrightarrow z)=0.
$$
Upon integrating by parts the above expression can be rewritten as
$$\int \textrm{d}^{3}z h(x)\left(2\nabla_{i_{1}}f\nabla_{i_{2}}(G^{abi_{1}i_{2}}\beta_{ab})+f\nabla_{i_{1}}\nabla_{i_{2}}(G^{abi_{1}i_{2}}\beta_{ab})\right)=0,$$
and since $f$ is arbitrary, the above condition implies that the terms within the round brackets ought to vanish which implies that:
\be \nabla_{i_{2}}(G^{abi_{1}i_{2}}\beta_{ab})=0. \label{divfree}\ee
From a theorem of Teitelboim,\footnote{In his unpublished PhD thesis. See \cite{Kuchar_Lagrangian} for a sketch of the argument.} given any functional $W^{ab}$ satisfying the divergence-free condition, \eqref{divfree}, $\nabla_a W^{ab}=0$  and the functional curl-free condition, \eqref{curlfree}, 
$$\frac{\delta (W^{kl}(y))}{\delta g_{ij}(x)}-\frac{\delta (W^{ij}(x))}{\delta g_{kl}(y)}=0 $$ it follows  that 
 $$W^{ab}=\frac{\delta c[g]}{\delta g_{cd}(x)}$$
 for some functional of the metric and its derivatives $c[g]$. Thus: 
\be \beta_{ab}(x)=G_{abcd}(x)\frac{\delta c[g]}{\delta g_{cd}(x)},\ee

In fact, we no longer need to consider higher order parts of the momentum independent Poisson bracket and the $\mathcal{O}(\pi)$ bracket.  To see the reason why,  note that the Hamiltonian constraint can be re-written as
\begin{align}&\left(G_{abcd}\pi^{ab}\pi^{cd}-\frac{\delta c[g]}{\delta g_{ab}}G_{abcd}\pi^{cd}\right)+\sqrt{g}\hat{V}[g;x)=\nonumber \\
&G_{abcd}\left(\pi^{ab}-\frac{\delta c[g]}{\delta g_{ab}}\right)\left(\pi^{cd}-\frac{\delta c[g]}{\delta g_{cd}}\right)-G_{abcd}\frac{\delta c[g]}{\delta g_{ab}}\frac{\delta c[g]}{\delta g_{cd}}+\sqrt{g}\hat{V}[g;x)\end{align}
The reason we re-wrote the Hamiltonian constraint in the above form is because we can perform a canonical transformation to redefine $\pi^{ab}-\frac{\delta c[g]}{\delta g_{ab}}\rightarrow \pi^{ab}$. Then the kinetic term is of the standard form and the potential gets redefined as 
$$\sqrt{g}\hat{V}[g;x)-G_{abcd}\frac{\delta c[g]}{\delta g_{ab}}\frac{\delta c[g]}{\delta g_{cd}}\equiv \sqrt{g}U[g;x).$$
The re-defined Hamiltonian constraint then reads
\be H[g,\pi;x)=G_{abcd}\pi^{ab}\pi^{cd}+\sqrt{g}U[g;x).\ee

Now we see that the task of computing the bracket $\mathcal{O}(\pi)$ term resulting from the Poisson bracket we were considering is recast into  computing the Poisson bracket between the new ultra-local kinetic term and the potential. There is no need to include any  term linear in the momentum, as the closure relations demand that it solely the result of canonical transformation. We thus choose the phase space coordinates where such a term is not present.  But this is the case that Farkas and Martinec  considered \cite{Martinec}, namely one where we have an ultra local kinetic term and an arbitrary scalar function of the metric and it's derivatives as the potential. In their analysis they prove that in such a scenario, the closure requirement for the Hamiltonian constraint will force the potential to take the form of that of General Relativity. This means that 
$$U[g;x)=\hat{V}[g;x)-\frac{1}{\sqrt{g}}G^{abcd}\beta_{ab}(x)\beta_{cd}(x)=R-2 \Lambda,$$
which is the potential term of the Hamiltonian constraint in General Relativity. 

   \section{Outlook}
   
  In this paper, we have for the first time extended the rigidity of the gravitational  Hamiltonian constraint in metric variables to all (finite) orders in spatial derivatives, for all terms. Our limitation was that we restricted ourselves to terms that were at most quadratic in momenta. We would however like to cover a couple of possible caveats to the full extension of the results to all orders. Before discussing terms with higher powers of the momenta, we will first comment on how the proofs we have presented extend to higher spatial dimensions. 
  
  \subsection{Extending the proofs to higher dimensions}
  
  The only place where we've used the fact that the spatial dimension is three is in assuming that the structures $\mathcal{B}^{i_{1}\cdots i_{m} j_{1}\cdots j_{n}}_{abcd}(g)$ and $\beta_{ab}(g)$ depends only on the Ricci tensor and its derivatives. \ \
  \paragraph*{The kinetic term}
  \ \
  
  The first place in the proof where this fact has been used was in our analysis of the terms containing the highest power of the covariant derivative operator. As a reminder, one needs to consider $\left(\left\{F(f),F_{o}(h)\right\}-(f\leftrightarrow h)\right)|_{\nabla^{(max)},h(x)=\delta(x)}$, which contains the expression 
  
   \begin{eqnarray}
\sum^{r'}_{r=0}\Bigg\{\nabla_{l_{2}}\nabla_{l_{1}}\nabla_{a_{r}}\cdots\nabla_{a_{1}}\left({f}\frac{\partial\left(\mathcal{B}^{i_{1}\cdots i_{n}j_{1}\cdots j_{m}}_{abcd}\nabla_{i_{1}}\cdots \nabla_{i_{n}}\pi^{cd}\nabla_{j_{1}}\cdots \nabla_{j_{m}}\pi^{ab}\right)}{\partial(\nabla_{a_{1}}\nabla_{a_{2}}\cdots\nabla_{a_{r}}R^{h}_{efg})}\right)\left(\,^{(0)}\mathcal{F}_{efg}^{hl_{1}ijl_{2}}G_{ijuv}\pi^{uv}\right)\label{rr1}\\
-(-1)^{r}\left(\frac{\partial\left(\mathcal{B}^{i_{1}\cdots i_{n}j_{1}\cdots j_{m}}_{abcd}\nabla_{i_{1}}\cdots \nabla_{i_{n}}\pi^{cd}\nabla_{j_{1}}\cdots \nabla_{j_{m}}\pi^{ab}\right)}{\partial(\nabla_{a_{1}}\nabla_{a_{2}}\cdots\nabla_{a_{r}}R^{h}_{efg})}\right) \nabla_{a_1}\cdots\nabla_{a_{r}}\nabla_{l_{1}}\nabla_{l_{2}}\left(f(x)\,^{(0)}\mathcal{F}_{efg}^{hl_{1}ijl_{2}}G_{ijuv}\pi^{uv}\right)\Bigg\}.\label{rr2}
\end{eqnarray}
Here, $\,^{(0)}\mathcal{F}_{efg}^{hl_{1}ijl_{2}}=g^{hp}\Xi^{l_{1}ij}_{pe[f}\delta^{l_{2}}_{g]}$. This structure comes from noticing that $\delta R^{h}_{efg}=\nabla_{f} \delta\Gamma^{h}_{ge}-\nabla_{g}\delta \Gamma^{h}_{fe}$ and then writing this in terms of the variation of the metric: $\nabla_{f} \delta\Gamma^{h}_{ge}-\nabla_{g}\delta \Gamma^{h}_{fe}=\,^{(0)}\mathcal{F}_{efg}^{hl_{1}ijl_{2}}\nabla_{l_{1}}\nabla_{l_{2}}\delta g_{ij}$. 
Now, here too, we choose the phase space function 
$$ f\nabla_{l_{2}}\nabla_{l_{1}}\nabla_{a_{r'-s}}\cdots\nabla_{a_{1}}\left(\nabla_{i_{1}}\cdots \nabla_{i_{n}}\pi^{cd}\nabla_{j_{1}}\cdots \nabla_{j_{m}}\pi^{ab}\right),$$

for $0\leq s\leq r'$, whose coefficient ought to strongly vanish in order for the Hamiltonian constraint to remain first class. This implies the vanishing of the set of phase space functions
\begin{equation*} \frac{\partial \mathcal{B}^{i_{1}\cdots i_{n}j_{1}\cdots j_{m}}_{abcd}}{\partial(\nabla_{a_{1}}\nabla_{a_{2}}\cdots\nabla_{a_{r'-s}}R^{h}_{efg})}\left(\,^{(0)}\mathcal{F}_{efg}^{hl_{1}ijl_{2}}G_{ijuv}\pi^{uv}\right), \end{equation*}
again for $0\leq s\leq r'$. We see that 
$$\,^{(0)}\mathcal{F}_{efg}^{hl_{1}ijl_{2}}G_{ijuv}\pi^{uv}=\textrm{tr}\pi\left(\frac{1}{2}\delta^{l_{2}}_{e}\delta^{l_{1}}_{[f}\delta^{h}_{g]}+\frac{1}{2}\delta^{h}_{e}\delta^{l_{1}}_{[f}\delta^{l_{2}}_{g]}+\frac{1}{2}g^{hl_{2}}g_{e[f}\delta^{l_{1}}_{g]}\right)+\pi^{h}_{e}\delta^{l_{1}}_{[f}\delta^{l_{2}}_{g]}+\pi^{h}_{[f}\delta^{l_{1}}_{g]}\delta^{l_{2}}_{e}+\delta^{l_{1}}_{[f}\pi_{g]e}g^{h l_{2}},$$
and is thus some non vanishing function of the momenta and so we conclude like before that the requirement that the Hamiltonian constraint remains first class demands that 
$$
\frac{\partial \mathcal{B}^{i_{1}\cdots i_{n}j_{1}\cdots j_{m}}_{abcd}}{\partial(\nabla_{a_{1}}\nabla_{a_{2}}\cdots\nabla_{a_{r'-s}}R^{h}_{efg})}=0,
$$
again, for the same range of $s$. Similarly, we arrive at the same conclusions as in the (3+1) dimensional case for $m=0$ (or $n=0$) where we choose the phase space functions $ f\nabla_{l_{2}}\nabla_{l_{1}}\nabla_{a_{r'-s}}\cdots\nabla_{a_{1}}\left(\nabla_{i_{1}}\cdots \nabla_{i_{n}}\pi^{cd}\pi^{ab}\right)$ and the case when $m=n=0$ where we choose the phase space functions $\nabla_{l_{2}}\nabla_{l_{1}}\nabla_{a_{r'-s}}\cdots\nabla_{a_{1}}\left(\pi^{cd}\pi^{ab}\right)$ and then requiring that their coefficients vanish will lead to the vanishing of $\frac{\partial \mathcal{B}^{i_{1}\cdots i_{n}}}{ \partial(\nabla_{a_{1}}\nabla_{a_{2}}\cdots\nabla_{a_{r'-s}}R^{h}_{efg})}$ and $\frac{\partial \mathcal{B}}{ \partial(\nabla_{a_{1}}\nabla_{a_{2}}\cdots\nabla_{a_{r'-s}}R^{h}_{efg})}$ for $0\leq s \leq r'$. 
  
  \paragraph*{The Linear in momentum term} \ \

The second place we used the fact that the the Riemann tensor is written entirely in terms of the Ricci tensor is when dealing with the term linear in the momenta, i.e. $P[g,\pi;x)=\beta_{ab}[g;x)\pi^{ab}(x)$ and considering its Poisson bracket with the potential: $\left\{P[g,\pi;x),\sqrt{g}\hat{V}[g;y)\right\}-(x\leftrightarrow y)$ that is independent of the momenta. We will consider the lowest order in spatial derivative terms in this bracket. This takes the form
$$\left(\frac{\partial \hat{V}[g,x)}{\partial R^{h}_{efg}}\right)\Bigg|_{(\partial^{(min)}g)}\,^{(0)}\mathcal{F}_{efg}^{hl_{1}ijl_{2}}\beta_{ij}(x)\nabla^{x}_{l_{1}}\nabla^{x}_{l_{2}}\delta(x,y).$$
Then, upon noting that $\left(\frac{\partial \hat{V}[g,x)}{\partial R^{h}_{efg}}\right)\Bigg|_{(\partial^{(min)}g)}=\delta^{f}_{h}g^{eg}$ so that $\delta^{f}_{h}g^{eg}\,^{(0)}\mathcal{F}_{efg}^{hl_{1}ijl_{2}}=G^{ijl_{1}l_{2}}$ and then smearing the above expressions with an arbitrary smearing function like we did in the three dimensional case that we find
$$\nabla_{l_{1}}(G^{ijl_{l}l_{2}}\beta_{ij})=0,$$
as the condition necessary for the vanishing of the above bracket lowest order in spatial derivatives. 

   Thus we find no obstruction to extending these proofs to higher dimensions, except we have still maintained the fact that the kinetic term is quadratic in the momenta. In higher dimensions, the existence of the Lovelock theories of gravity which still allow for second order equations of motion begs the question why our analysis seems to rule them out. As we will see in the coming subsection, this is because such theories necessarily require higher powers of the momentum to appear in the Hamiltonian constraint. In the following subsection we will briefly sketch how one obtains a canonical Hamiltonian in such theories. 
  \subsection{The Hamiltonian Constraint of Lovelock gravity}
Lovelock theories of gravity in dimensions higher than four are, like General Relativity, theories of a dynamical metric and whose Hamiltonian is a sum of $D$ first class constraints and one scalar constraint (so the dimension of the space-time is $D+1$).  In this sub-section we will closely follow the analysis of \cite{Teitelboim_Zanelli, Liu_Sabra}. The Lagrangian density for such theories reads

$$\mathcal{L}=\pi^{ab}\dot{g}_{ab}-NH-N^{a}H_{a},$$

We can now define the extrinsic curvature tensor as $K_{ab}=\frac{1}{2N}(\dot{g}_{ab}-\nabla_{(a}N_{b)})$, and then we find that the diffeomorphism constraint can be absorbed into the symplectic term and the Lagrangian density can re-written as
$$\mathcal{L}=\pi^{ab}K_{ab}-NH.$$
where now the Hamiltonian is a function of the configuration variables,  $H(g_{ab}, K_{ab})$. 
There exists a relationship between the extrinsic curvature and the momentum which can be solved for from the relation

$$\pi_{ab}=\frac{\partial \mathcal{L}}{\partial \dot{g}_{ab}}.$$
This relation is simple, linear, and thus invertible in General Relativity. 

The Lagrangian density of Lovelock gravity is given by the double sum

\be \mathcal{L}=\sum_{2p<D+1}\sqrt{g}\alpha_{p} \sum^{p}_{s=0}C_{s(p)}\delta^{[i_{1}\cdots i_{2s}\cdots i_{2p}]}_{[j_{1}\cdots j_{2s}\cdots j_{2p}]}R^{j_{1}j_{2}}_{\,\,\,i_{1}i_{2}}\cdots R^{j_{2s-1}j_{2s}}_{\,\,\,i_{2s-1}i_{2s}} K^{j_{2s+1}}_{i_{2s+1}}\cdots  K^{j_{2p}}_{i_{2p}}.\ee
Some explanation of the structures given is necessary: $\delta^{[i_{1}\cdots i_{2s}\cdots i_{2p}]}_{[j_{1}\cdots j_{2s}\cdots j_{2p}]}$ is the totally anti-symmetrised product of $2p$ Kronecker symbols except there is no factor of $\frac{1}{(2p)!}$ for the upper indices' anti-symmetrisation, $\left\{\alpha_{p}\right\}$ can be considered the set of coupling constants, and $C_{s(p)}$ is a combinatorial factor which reads

$$C_{s(p)}=\frac{(-4)^{p-s}}{s![2(p-s)-1]!!}.$$

The momentum coming from the partial derivative of the Lagrangian density with respect to the metric velocity term can be written as a sum:

$$ \pi^{i}_{j}=\sum_{p}\alpha_{(p)}\pi^{i}_{j(p)},$$
where each term of the sum encodes the relation between the momenta and the extrinsic curvature reads
\be \pi^{i}_{j(p)}= -\frac{1}{4}\sqrt{g}\sum^{p-1}_{s=0}C_{s(p)}\delta^{[i_{1}\cdots i_{2s}\cdots i_{2p-1}i]}_{[j_{1}\cdots j_{2s}\cdots j_{2p-1}j]}R^{j_{1}j_{2}}_{\,\,\,i_{1}i_{2}}\cdots R^{j_{2s-1}j_{2s}}_{\,\,\,i_{2s-1}i_{2s}} K^{j_{2s+1}}_{i_{2s+1}}\cdots  K^{j_{2p-1}}_{i_{2p-1}}.\ee
 Now, the inversion of the above relation cannot be done in closed form. Suppose however that we set $\alpha_{0}=\alpha_{1}=1$ and then call $\alpha_{2}=\alpha$ and choose to truncate the expansion of the Lagrangian to just the first three terms. Then we are left with what is known as Gauss-Bonnet Gravity. 

The second order is also the first non trivial order beyond the Lagrangian of General Relativity. This higher derivative term in four dimensions is topological and thus unseen by the dynamics, but in five and higher dimensions, it too contributes non trivially to the dynamics.  The Hamiltonian constraint in this case can be re-written in term of the canonical momenta. The Gauss--Bonnet terms are the first non trivial higher derivative Lovelock terms beyond General Relativity, and this fact will also be apparent in the Hamiltonian formulation of the theory. Now, if the truncation in the sum over $p$ in the canonical momentum defined above happens at second order, i.e. $\pi^{ab}=\pi_{(1)}^{ab}+\pi^{ab}_{(2)},$ then we find 

$$\pi^{ab}_{(2)}=-2\bigg(g^{ab}(R\textrm{tr}K-2R_{cd}K^{cd})-RK^{ab}-2R^{ab}\textrm{tr}K+4R_{c}^{(a}K^{b)c}+2R^{acbd}K_{cd}+$$ $$\frac{1}{3}g^{ab}(-\textrm{tr}K^{3}+3KK^{cd}K_{cd}-2K^{d}_{c}K^{e}_{d}K^{c}_{e})+\textrm{tr}K^{2}K^{ab}-2\textrm{tr}KK^{a}_{c}K^{bc}-K^{ab}K^{cd}K_{cd}+2K^{a}_{c}K^{c}_{d}K^{bd}\bigg).$$

We can exploit the separation of scales between $\alpha_{1}$ and $\alpha_{2}$ to write $\pi^{ab}=K^{ab}-g^{ab}\textrm{tr}K+\mathcal{O}(\alpha)$ to lowest order, and then plug this back into the above expression to rewrite $\pi^{ab}_{(2)}$ in terms of this $\pi^{ab}$ to $\mathcal{O}(\epsilon^{2})$ and then write the Hamiltonian constraint to this order in terms of this momentum. 

The Hamiltonian constraint now takes the form 
$$ H[g,\pi;x)=\left(\sqrt{g}(R+\Lambda) +\frac{1}{\sqrt{g}}\bigg(\pi^{ij}\pi_{ij}-\frac{1}{D-2}\pi^{2}\bigg)\right)+\alpha\bigg(\sqrt{g}\bigg(R^{2}-4R_{ab}R^{ab}+R_{abcd}R^{abcd}\bigg)+ $$
$$\frac{1}{\sqrt{g}}\bigg(-\frac{16}{D-2}R_{ab}\pi \pi^{ab}+\frac{2D}{(D-2)^{2}}R\textrm{tr}\pi^{2}-2R\pi_{ab}\pi^{ab}+8R_{ab}\pi^{b}_{c}\pi^{ca}+4R_{abcd}\pi^{ac}\pi^{bd}\bigg)+$$
\be \frac{1}{g}\bigg(2\pi^{b}_{a}\pi^{c}_{b}\pi^{d}_{c}\pi^{a}_{d}-(\pi^{ab}\pi_{ab})^{2}-\frac{16}{3(D-2)}\pi \pi^{b}_{a}\pi^{c}_{b}\pi^{a}_{c}+\frac{2D}{(D-2)^{2}}\pi^{2}\pi^{ab}\pi_{ab}-\frac{3D-4}{3(D-2)^{3}}\pi^{4}\bigg)\bigg)+\mathcal{O}(\alpha^{2})\ee

If we didn't truncate to just the Gauss--Bonnet terms and retain only terms linear in $\alpha$, then we wouldn't have been able to obtain the inversion required to write the extrinsic curvature tensor in terms of the canonical momentum and thereby obtain the Hamiltonian as a function of the metric and the momenta. Nevertheless, it is apparent why our previous analysis couldn't have seen these Lovelock theories in higher dimensions: the linear in $\alpha$ corrections to the Hamiltonian constraint of general relativity contain up to four powers of the momenta, in addition to what appears as corrections to the kinetic term which are no longer ultra local in the metric and the momenta. Notice also that the quartic term is ultra local in the metric and the momenta and furthermore,  like General Relativity,  Lovelock theory too is time-reversal symmetric and hence contains only even powers of the momenta. 

It would indeed be interesting to fully understand whether analysis akin to ours can be extended beyond the restriction of two powers of the momenta in the Hamiltonian constraint. Yet answering this question lies far beyond the scope of this article. What is interesting to note is that the restriction to two powers of the momenta in the Hamiltonian constraint seems to be the condition that singles out General Relativity among other space-time covariant theories of gravity that contain only two time derivatives in equations of motion.\footnote{ Moreover, there are known possible issues for dynamical systems containing higher order of momentum, known as Ostrogadsky\rq{}s instabilities, which this simple demand avoids. } 

\section*{Acknowledgments}
We would like to thank Sung-Sik Lee for constant support in regards to this project, and for pointing out multiple loopholes in various stages of it. We would also like to thank Lee Smolin for constant support. 

This research was supported in part by Perimeter Institute for Theoretical Physics as well as by grant from NSERC and the John Templeton Foundation. Research at
Perimeter Institute is supported by the Government of Canada through the Department of Innovation,
Science and Economic Development Canada and by the Province of Ontario through the Ministry of
Research, Innovation and Science.


\end{document}